\documentclass[aps,prl,groupedaddress,twocolumn,superscriptaddress]{revtex4-1}

\usepackage{graphicx}
\usepackage{textgreek}

\usepackage{multirow}
\usepackage{braket}
\usepackage{amssymb}

\begin{document}

\title{High-precision mass spectrometer for light ions}
\author{F. Hei\ss e}
\email{This article contains material from the PhD work of F. Hei\ss e and S. Rau, both  enrolled at the Ruprecht-Karls University of Heidelberg.}
\affiliation{Max-Planck-Institut f\"ur Kernphysik, Saupfercheckweg 1, 69117 Heidelberg, Germany}
\affiliation{GSI Helmholtzzentrum f\"ur Schwerionenforschung GmbH, Planckstra\ss e 1, 64291 Darmstadt, Germany}
\author{S. Rau }
\affiliation{Max-Planck-Institut f\"ur Kernphysik, Saupfercheckweg 1, 69117 Heidelberg, Germany}
\author{F. K\"ohler-Langes}
\affiliation{Max-Planck-Institut f\"ur Kernphysik, Saupfercheckweg 1, 69117 Heidelberg, Germany}
\author{W. Quint}
\affiliation{GSI Helmholtzzentrum f\"ur Schwerionenforschung GmbH, Planckstra\ss e 1, 64291 Darmstadt, Germany}
\author{G. Werth}
\affiliation{Institut f\"ur Physik, Johannes Gutenberg-Universit\"at, 55099 Mainz, Germany}
\author{S. Sturm}
\affiliation{Max-Planck-Institut f\"ur Kernphysik, Saupfercheckweg 1, 69117 Heidelberg, Germany}
\author{K. Blaum}
\affiliation{Max-Planck-Institut f\"ur Kernphysik, Saupfercheckweg 1, 69117 Heidelberg, Germany}

\date{\today}

\begin{abstract}
 
The precise knowledge of the atomic masses of light atomic nuclei, e.g. the proton, deuteron, triton and helion, is of great importance for several fundamental tests in physics. However, the latest high-precision measurements of these masses carried out at different mass spectrometers indicate an inconsistency of five standard deviations. To determine the masses of the lightest ions with a relative precision of a few parts per trillion and investigate this mass problem a cryogenic multi-Penning trap setup, $\text{LIONTRAP}$ (Light ION TRAP), was constructed. This allows an independent and more precise determination of the relevant atomic masses by measuring the cyclotron frequency of single trapped ions in comparison to that of a single carbon ion. In this paper the measurement concept and the first doubly compensated cylindrical electrode Penning trap, are presented. Moreover, the analysis of the first measurement campaigns of the proton's and oxygen's atomic mass is described in detail, resulting in $m_{\text{p}} = 1.007\,276\,466\,598\,(33) \, \text{u}$ and $m\left(^{16}\text{O}\right) = 15.994\,914\,619\,37\,(87)\,\text{u}$. The results on these data sets have already been presented in [F. Hei\ss e et al., Phys. Rev. Lett. 119, 033001 (2017)]. For the proton's atomic mass, the uncertainty was improved by a factor of three compared to the 2014 CODATA value.

\end{abstract}

\maketitle

\section{Introduction} \label{intro_1}

The Standard Model of particle physics (SM) compiles our current state of knowledge on fundamental physics. With its help, it is possible to precisely calculate observables in a wide range of fields. Experiments that measure these derived observables can thus be used to probe the validity of the SM. Despite its beauty and potency, the theories contained within SM however require the knowledge of a considerable number of so-called fundamental constants in order to allow predictions. Especially the field of atomic and molecular physics has originated a wealth of intriguing experiments which allow the search for deviations from known physics with ever-increasing resolution, and which consequently require the knowledge of the fundamental constants with previously inaccessible precision~\cite{RevModPhys.90.025008}. 
Prominent among these constants are the rest masses of fundamental particles such as the electron, but also those of composite particles such as the proton or the neutron and generally the lightest elements~\cite{RevModPhys.88.035009,Mohr_2018}. A recent review about the masses of these lightest elements can be found in~\cite{Atoms2019.7.37}. In Table~\ref{top_5} the most precise determinations of those masses have been compiled.    

\begin{table} 
\begin{center}
\caption{Overview of the most precise masses directly referenced to $^{12}\text{C}$ and their uncertainties. All values in the table correspond to the individual mass measurement of this particle with the lowest uncertainty. The majority of the values have been determined in the UW-PTMS experiment (University of Washington Penning-Trap Mass Spectrometer) by Van Dyck Jr. at the University of Washington (UW), whereas the others have been measured by the authors' group (MPIK).}\label{top_5}
\begin{ruledtabular}
\begin{tabular}{l r c c}
Particle & Atomic masses (u) &  $\frac{\delta m}{m} (10^{-12})$  & Group \\
 \hline
 e$^-$  & $0.000\,548\,579\,909\,069\,(15)$ & 28 & MPIK~\cite{Nature2014,0953-4075-48-14-144032}\footnote{The atomic mass of the electron is determined by combining a high-precision measurement of the Larmor-to-cyclotron frequ- ency ratio of $^{12}\text{C}^{5+}$ with bound state quantum electrodynamics calculations of its \textit{g}-factor.} \\
 p$^+$ & $1.007\,276\,466\,598\,(33)$ & 33 &  MPIK~\cite{PhysRevLett.119.033001}\footnote{The value varies by $15\times 10^{-12}\,\text{u}$ in comparison to the one reported in the cited source. This is due to a shift of the temperature discovered in the reanalysis, which is described in sections~\ref{chap_MT} and \ref{mp_stat}.} \\
d$^+$ & $2.013\,553\,212\,745\,(40)$ & 20 & UW~\cite{0026-1394-52-2-280} \\
$^3$He & $3.016\,029\,321\,675\,(43)$& 14 & UW~\cite{0026-1394-52-2-280} \\
$^4$He  & $4.002\,603\,254\,131\,(62)$ &  15 & UW~\cite{PhysRevLett.92.220802,VANDYCK2006231} \\
$^{16}\text{O}$ & $15.994\,914\,619\,57(18)$  & 11 & UW~\cite{VanDyck2001,VANDYCK2006231} \\
\end{tabular}
\end{ruledtabular}
\end{center}
\end{table}
Such masses are generally measured in Penning-trap experiments, where the ratio (CFR) of the cyclotron frequencies $\nu_c=\frac{q}{2 \pi m} B$ of two ions with charge $q$ is measured within the same magnetic field $B$, allowing to relate the masses $m$ of the ion of interest to that of a known reference mass. This reference mass could be an ion of $^{12}$C, in which case the mass of interest can be directly related to the atomic mass unit $u$. Today, a network of CFRs allows indirectly connecting measured mass ratios with the atomic mass unit. An analysis and compilation of all known atomic masses is provided by the team of the Atomic Mass Evaluation (AME)~\cite{1674-1137-41-3-030002,Wang_2017}.

Unfortunately, especially the measurement of the interesting light ion (and particle) masses are complicated by the sizable systematic frequency shifts originating in the relatively large ratio of kinetic energies compared to the low rest mass energy. 
Consequently, we have developed the LIONTRAP (Light Ion TRAP) apparatus, which is optimized to minimize these systematic shifts. Recently, as a first application, we have performed a measurement of the proton mass, where we have achieved a relative precision of $33$ parts per trillion (ppt)~\cite{PhysRevLett.119.033001}. Combined with the electron atomic mass~\cite{Nature2014,0953-4075-48-14-144032}, previously measured in our group, the proton mass enters the Rydberg constant $R_\infty$ via the reduced mass $\mu$ of the hydrogen atom~\cite{RevModPhys.88.035009,Mohr_2018,Karshenboim2016}. As many of the most precisely measured CFRs involve molecules containing one or several hydrogen atoms, the proton mass is also linked to other masses, such as $^{13}$C, $^{15}$N, $^{29}$Si, $^{31}$P and $^{33}$S~\cite{1674-1137-41-3-030002,Wang_2017}.

\begin{figure}
\includegraphics[width=0.483\textwidth]{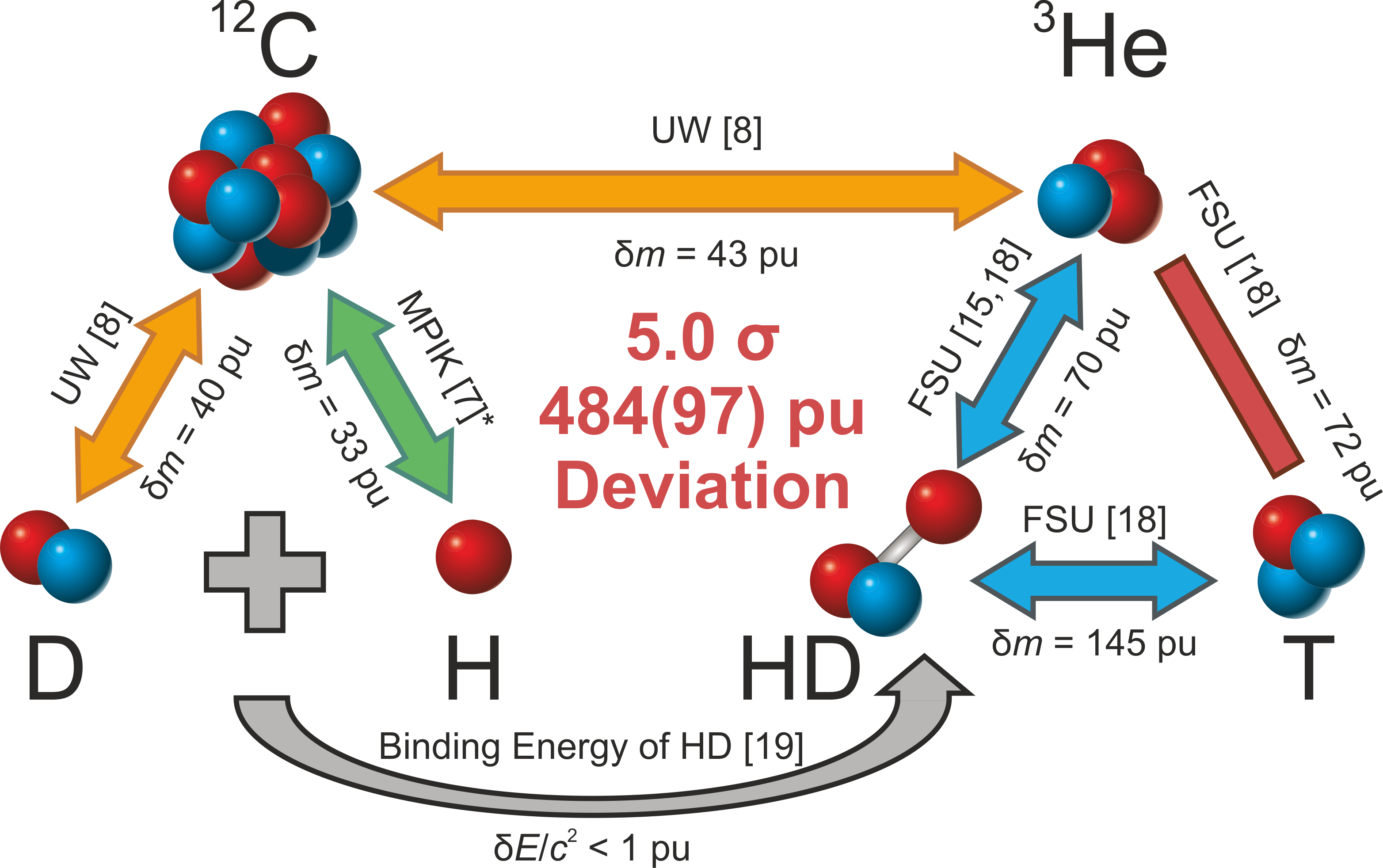}
\caption{\label{puzzle} The puzzle of light ion masses. The orange and blue links are CFRs measured at the UW-PTMS~\cite{0026-1394-52-2-280} and the group led by Edmund Myers at the Florida State University in Tallahassee (FSU)~\cite{PhysRevA.96.060501,PhysRevLett.114.013003}, respectively. The green link is the proton's atomic mass measured by the $\text{LIONTRAP}$ experiment. The mass after the reanalysis, given in this article, is used (indicated by the asterisk, for details see sections~\ref{chap_MT} and \ref{mp_stat}). Here, only the most precise measurements together with their absolute uncertainties in pu ($10^{-12}\,\text{u}$) are shown for each link. Since the mass of the HD molecule can be calculated from the masses of D and H together with its binding energy~\cite{PhysRevA.67.062504}, this gives an additional link. A $5.0\,\sigma$ discrepancy remains by applying all links. Furthermore, the red bar shows the required CFR of $\text{KATRIN}$.}
\end{figure}

However, our value deviates by about three standard deviations from the one previously tabulated by CODATA~\cite{RevModPhys.88.035009}. This discrepancy is part of a broader problem in the light mass range (see Fig.~\ref{puzzle}), involving the masses of proton, deuteron, triton ($^{3}$H$^{+}$) and the helion ($^{3}$He$^{2+}$)~\cite{PhysRevA.96.060501}. The ratios of these masses, measured by different groups, are currently inconsistent with each other by about five standard deviations. This discrepancy needs to be resolved in order to restore trust in the value for the mass difference of triton and helion, which in turn is required to extract or bound the electron antineutrino rest mass with the KATRIN experiment~\cite{KATRINCollaborationKATRINCollaboration2005_270060419,Otten2010}.

With the Penning-trap experiment $\text{LIONTRAP}$ we are measuring the masses of the lightest ions in atomic mass units, starting with the proton. In this article we discuss the measurement concept and the systematic uncertainty budget of LIONTRAP in more detail than in the original Letter. This article is structured as follows: The measurement principles and the detection techniques of $\text{LIONTRAP}$  as well as the setup are described in section~\ref{det_tec}. In section~\ref{chap_MT} our doubly-compensated Penning trap is introduced, including the determination of the magnetic field inhomogeneity and the temperature of the particles. In sections~\ref{mp_stat} and \ref{mo_stat} a detailed evaluation of the proton's and oxygen's atomic mass is presented. An outlook and some final remarks conclude this article (section~\ref{outlook}).

\section{Experimental background} \label{det_tec}

\begin{flushleft}
\textbf{Measurement fundamentals}
\end{flushleft}

The basics of Penning trap physics have been described in~\cite{PENNING1936873,pierce1954,RevModPhys.58.233,RevModPhys.62.525}. To achieve a consistent notation, the most important formulas are reviewed here. A Penning trap consists of a homogeneous static magnetic field, which is superimposed with an electrostatic quadrupole potential to confine the ion in the $z$-direction, leading to an electrostatic potential $V$ in cylindrical coordinates ($z$,$\rho$):
\begin{equation}
V(z,\rho)=\frac{U_{\text{R}}}{2}  \sum_{n=0,2,4}^{\infty} \frac{ C_n }{d^n_{char}} \sum_{k=0}^{\frac{n}{2}}\frac{\left(-1\right)^k n! \cdot z^{n-2k} \rho^{2k}}{2^{2k} \left(n-2k\right)!  \left(k!\right)^2} \, , \label{pot_tay}
\end{equation}
where $U_{\text{R}}$ is the voltage of the inner electrode (ring electrode), $d_{\text{char}}=\sqrt{(2\cdot d_0^2+r_0^2)/4} $ is a characteristic trap size, defined by the trap radius $r_0$ and the axial distance of the endcap from the trap center $d_0$. The ideal harmonic trapping potential is defined by: $C_4=C_6=~\ldots~=~C_{\infty}=0$. The precision trap of the $\text{LIONTRAP}$ experiment is designed to have $C_2=-0.5997$ and $d_{\text{char}}= 5.107\,\text{mm}$.

The combination of the two fields yields an ion motion which can be decomposed into three independent harmonic eigenmotions: two radial modes, the modified cyclotron motion with frequency $\nu_+$ and the magnetron motion with frequency $\nu_-$, as well as the axial motion with frequency $\nu_z$. For an ideal Penning trap the three eigenfrequencies can be expressed by:
\begin{eqnarray}
\nu_z &=& \frac{1}{2 \pi} \sqrt{\frac{q}{m}\frac{U_{\text{R}} C_2}{ d^2_{\text{char}}}} \quad , \\
\nu_+ &=& \frac{1}{2} \left(\nu_c + \sqrt{\nu_c^2-2\nu_z^2} \right) \quad ,\\
\nu_- &=& \frac{1}{2} \left(\nu_c - \sqrt{\nu_c^2-2\nu_z^2} \right) \quad .
\end{eqnarray}

Our experimental conditions are $\vec{B}=B_z \approx 3.8\,\text{T}$ and $U_{\text{R}} \approx -10\,\text{V}$. Especially for highly charged ions with $q/m \gtrsim 0.5\,$e/u there is a strong hierarchy: $\nu_+ \gg \nu_z \gg \nu_-$. The free cyclotron frequency of an ion can be determined via the invariance theorem~\cite{PhysRevA.25.2423}:
\begin{equation}
\nu_c = \sqrt{\nu_+^2 + \nu_z^2 + \nu_-^2} \quad .
\end{equation}
This formula is invariant with respect to a tilt between the axis of the trap electrodes and the magnetic field axis, as well as an ellipticity of the electric potential.

To calibrate the magnetic field we chose a bare carbon nucleus  $^{12}\text{C}^{6+}$ as a reference ion since its atomic mass can be determined with very small uncertainty:
\begin{equation}
m(^{12}\text{C}^{6+})= m(^{12}\text{C})-6  m_{\text{e}}+\sum_{i=1}^{6}\frac{E_{b,i}}{c^2} \quad . \label{eq:2}
\end{equation}

\begin{table} 
\begin{center}
\caption{Overview of all electronic binding energies and their uncertainties of the carbon atom $^{12}\text{C}$. The conversion factor $e/c^2=1.073\,544\,110\,7 \times 10^{-9}\,\text{u/eV}$, based on CODATA 2014~\cite{RevModPhys.88.035009}, is applied to convert the given CODATA values into corresponding masses.}\label{binding_c}
\begin{ruledtabular}
\begin{tabular}{l c c}
Ionisation level & Binding energy $E_{\text{b}} \left(\text{eV}\right)$  & Reference \\
 \hline
1$^+$ & $11.260\,288\,(11)$ & \cite{0067-0049-233-1-16} \\
2$^+$ & $24.384\,5\,(9)$ & \cite{BIEMONT1999117} \\
3$^+$ & $47.887\,78\,(12)$  & \cite{1402-4896-1-5-6-013} \\
4$^+$ & $64.493\,52\,(19)$ & \cite{1402-4896-55-6-010} \\
5$^+$ & $392.090\,515\,(25)$ & \cite{doi:10.1139/p88-100} \\
6$^+$ & $489.993\,194\,(7)$ & \cite{JOHNSON1985405} \\
\end{tabular}
\end{ruledtabular}
\end{center}
\end{table}
Here, $E_{b,i}$ are the binding energies of the six removed electrons of the carbon atom, see Table~\ref{binding_c}. The mass $m(^{12}\text{C}^{6+})$ can be derived with a relative uncertainty of  $0.08\,\text{ppt}$ to $m(^{12}\text{C}^{6+})= 11.996\,709\,626\,413\,85(8)\,\text{u}$. This uncertainty arises mostly from the uncertainty of the second ionization level and does not limit the atomic mass measurement of the proton. In the case of the proton its atomic mass is expressed by: 
\begin{eqnarray}
m_{\text{p}} &=&  \frac{1}{6} \frac{\nu_c\left(^{12}\text{C}^{6+}\right)}{\nu_c\left(\text{p}\right)} m\left(^{12}\text{C}^{6+}\right) \\
&=& \frac{1}{6} \text{CFR} \left(^{12}\text{C}^{6+}, \text{p}\right)  m\left(^{12}\text{C}^{6+}\right) \quad . \label{eq:3}
\end{eqnarray}

An overview of the CFR measurements for the different experiments is given in~\cite{MYERS2013107,Atoms2019.7.37}. In the $\text{LIONTRAP}$ experiment $\nu_c$ of the proton and the carbon ion are measured in the same precision trap (PT) in subsequent measurements at small energies with a minimum time in between. To guarantee the same position within the trap, the trapping potential $U_{\text{R}}$ is the same for both ions, which results in very different axial frequencies of $\nu_z(\text{p}) \approx \sqrt{2} \cdot \nu_z(^{12}\text{C}^{6+})$. Therefore, separate detection systems for both ions are required, which need to be tuned very precisely. Furthermore, two storage traps are connected to the PT to park one ion at a time. Additionally, the trap tower of the LIONTRAP experiment contains another trap, the magnetometer trap (MT), which is intended for simultaneous phase-sensitive measurements to be used in a later phase of LIONTRAP.

\begin{flushleft}
\textbf{Detection techniques}
\end{flushleft}

The measurement of the axial frequency is performed non-destructively via the ion's induced image currents $i_{\text{ind}}$ on the surface of one or more trap electrodes, typically correction electrodes. Applying the Shockley-Ramo theorem~\cite{1686997} the induced image current yields:
\begin{eqnarray}
i_{\text{ind}}(t) & = &\frac{q}{D_{\text{eff}}} \dot{z}(t) \quad , \\
D_{\text{eff}} &\equiv & \frac{U_{\text{el}}} {\frac{\partial U_{\text{el}}}  {\partial z} \vert_{\rho=0,z=0}}  \quad ,
\end{eqnarray} 
where $D_{\text{eff}}$ is the effective geometric electrode distance and $U_{\text{el}}$ the voltage of the electrode connected to the corresponding superconducting tank circuit. All other electrodes are at zero voltage. While above formulas are true for detecting the axial motion, one can also detect $\nu_{\pm}$ by using vertically split electrodes and the derivative of the potential in the radial direction. At our experiment $D_{\text{eff}}$ is between $5\,\text{mm}$ and $30 \,\text{mm}$  depending on the chosen electrodes. Capacitive coupling between neighboring electrodes can lead to a modification of $D_{\text{eff}}$ for the respective detection system.

The ion induces currents in the order of femtoampere into the trap electrodes. For ion frequencies equal to the resonance frequency of the superconducting tank circuit a large real part of the impedance, corresponding to an effective electric parallel resistance $R_{\text{p}}$, is favorable to convert the small induced image currents into detectable voltages. $R_{\text{p}} = 2 \pi Q L  \nu_{\text{res}} = Q / (2 \pi C  \nu_{\text{res}})$, where $Q$ is the quality factor, $L$ is the inductance, $C$ the capacitance and  $\nu_{\text{res}} = 1 / \left(2 \pi \sqrt{ L C} \right)$ representing the resonance frequency of the tank circuit in the short-coil limit due to our relatively large capacitances in the order of $10\,\text{pF}$. The inductance is governed by the number of windings of the coil and their geometry. The capacitance depends on the geometry of the coil and the resonator housing as well as the sum of all connected capacitances of the detection system and the electrodes. Furthermore, the respective frequency of the ion and the resonator should be in resonance ($\nu_{\text{res}} =    \nu_z$ or $\nu_{\text{res}} =  \nu_+ $), which sets some limitations on the combination of $L$ and $C$. 
The resulting voltage, typically in the order of nanovolt, is amplified directly in the $4\,\text{K}$-electronic section by ultra low-noise cryogenic amplifiers, which feature a current noise of $i_{\text{n}} (\text{Amp}) < 10\,\text{fA}/\sqrt{\text{Hz}} $ and a voltage noise of $u_{\text{n}}(\text{Amp}) \approx 400\,\text{pV}/\sqrt{\text{Hz}}$ at axial and modified cyclotron frequencies~\cite{Sven2012,10.1063/1.4967493}. Additional room temperature amplifiers further increase the signal level. Later, the spectra of the axial and modified cyclotron frequency signals are down-mixed into a range from $0\,\text{kHz}$ to $28\,\text{kHz}$ using a single-sideband mixer.

The individual time-domain traces, each $32\,\text{s}$ long, are Fourier-transformed and then averaged for up to $192\,\text{s}$ in the frequency domain. Finally, the eigenfrequency is extracted from a least-square regression using a lineshape model~\cite{Sven2012}.

The trap setup as well as the superconducting tank circuits are cooled to the temperature of liquid helium ($4\,\text{K}$). In resonance with the tank circuit, the axial motion of the ion is resistively cooled and thermalized within about a second with the $4\,\text{K}$ cold resonator~\cite{doi:10.1063/1.321602}. In thermal equilibrium with the tank circuit the ion shortens the thermal noise of the resonator at its axial frequency $\nu_z$. This ion signal is called a dip.

For the proton the full width at half maximum (FWHM) of the dip in the frequency power spectrum is $660\,\text{mHz}$ at $740\,\text{kHz}$ and for the carbon ion $1100\,\text{mHz}$ at $525\,\text{kHz}$. The two radial eigenfrequencies can be measured by a coupling to the axial motion. For example $\nu_+$ can be determined by a continuous wave quadrupole coupling with the ``red" axial-cyclotron sideband at $ \nu_{\text{rf+}} =\nu_+-\nu_z$~\cite{PhysRevA.41.312}. For the magnetron frequency the ``blue" axial-magnetron sideband at $\nu_{\text{rf-}}=\nu_-+\nu_z$ is driven. During this sideband coupling the axial motional amplitude is modulated and thus the axial dip splits into two dips $\nu_{\text{left}}$ and $~\nu_{\text{right}}$ of the so-called double-dip. Additionally, during the sideband coupling, the two radial modes thermalize with the axial resonator. Finally, the modified cyclotron frequency and the magnetron frequency can be determined via the avoided-crossing relation:
\begin{eqnarray}
\nu_+ &= \nu_{\text{rf+}} - \nu_z +\nu_{\text{left}}+\nu_{\text{right}} \quad ,\\
\nu_- &= \nu_{\text{rf-}} + \nu_z -\nu_{\text{left}}-\nu_{\text{right}} \quad .
\end{eqnarray}

The determination of the frequencies is performed at low kinetic energies corresponding to $T_z \approx 4\,\text{K}$, $T_+ = \nu_+/\nu_z \cdot T_z $ and $T_- = - \nu_-/\nu_z \cdot T_z  $~\cite{RevModPhys.58.233}. The eigenfrequencies, temperatures, energies $\left(E_z , E_+, E_- \right)$ and the axial amplitude $z_0$ as well as the two radii $r_+~\text{and}~ r_- $ of the proton and carbon ion in our experiment are summarized in Table~\ref{eigen_f}.

\begin{table} 
\begin{center}
\caption{Overview of typical eigenfrequencies, temperatures, energies and amplitudes for a thermalized proton and $^{12}\text{C}^{6+}$ ion at $U_{\text{R}} \approx -10\,\text{V}$ and $B_z \approx 3.8\,\text{T}$. The determination of the temperatures is discussed in detail in section~\ref{chap_MT} and Appendix~\ref{ion_temp_app}.}\label{eigen_f}
\begin{ruledtabular}
\begin{tabular}{l c c}
 & proton  & $^{12}\text{C}^{6+}$ \\
 \hline
$\nu_-$ (Hz) & 4771.0 &  4771.4 \\
$\nu_z$ (Hz) & $739\,873$ & $525\,141$ \\
$\nu_+$ (Hz)& $57\,379\,350$ & $28\,903\,993$  \\
$\nu_c$ (Hz) & $57\,384\, 120$ &  $28\,908\,764$ \\
\hline
$T_-$ (K) & -0.022 &  -0.058\\
$T_z$ (K) & 3.4 & 6.4\\
$T_+$ (K) & 260 & 350 \\
\hline
$E_-$ (meV) & $-1.9 \times 10^{-3}$ & $-5.0 \times 10^{-3}$\\
$E_z$ (meV) & 0.31 & 0.55 \\
$E_+$ (meV) & 23 & 30 \\
\hline
$r_-$ ($\mu$m) & 5.8 & 3.8 \\
$z_0$ ($\mu$m) &  51 & 29 \\
$r_+$ ($\mu$m) & 5.8 & 3.8  \\
\end{tabular}
\end{ruledtabular}
\end{center}
\end{table}

The dip as well as the double-dip spectrum are averaged noise spectra $u_n$ with an unfavorable scaling of the precision with the measurement time $ T_{\text{meas}}$ of $\delta u_{\text{n}} / u_{\text{n}} \propto 1/\sqrt{T_{\text{meas}}}$. The relative uncertainty of the determination of the free cyclotron frequency is about $2 \times 10^{-9}$, when using a single dip spectrum and the two corresponding double dip spectra taken before and after the dip. The combined measurement time for the axial and modified cyclotron frequency accumulates to seven minutes. 

In order to mitigate the adverse effects of the long measurement time as well as the uncertainty arising from the dip lineshape model, a phase-sensitive method, PnA (Pulse and Amplify), was established in our group to determine $\nu_+$~\cite{PhysRevLett.107.143003}. There, the phase of the modified cyclotron motion is transferred to the axial motion via a short sideband coupling at the ``blue" sideband, $\nu_+ + \nu_z$, which transfers the cyclotron phase into the axial phase and amplifies both modes. Finally, the time signal of the excited axial motion (an axial peak signal) is read out to extract the axial phase. The PnA method results in an approximately one order of magnitude more precise modified cyclotron frequency determination than the double-dip method, as the relative precision of $\nu_+$, using the PnA method, scales favorably with $1/T_{\text{meas}}$ and is very fast ($\sim 10~\text{s}$). Consequently, the impact of the magnetic field fluctuations during the CFR measurement is significantly reduced. Compared to other phase-sensitive techniques, the PnA method works at small excitation energies resulting in correspondingly small systematic shifts. Moreover, it can be extrapolated to zero excitation energies of $\nu_+$, which further reduces energy dependent systematic shifts. Unlike the double-dip technique, PnA does not rely on a determination of the axial frequency in leading order and is thus significantly less prone to systematic uncertainties associated with the dip lineshape model.

For a reliable determination of the $\nu_+$ phase, a sufficiently large signal-to-noise ratio of the peak signal $\text{SNR}_{\text{peak}}$ is required to reduce the technical phase readout jitter~\cite{0953-4075-48-14-144032}. The $\text{SNR}_{\text{peak}}$ is proportional to the charge and the axial motional amplitude $z_0$ of the ion. Therefore, a stable phase-sensitive measurement of $\nu_+$ of a single proton is challenging due to its low charge. To guarantee a sufficiently high $\text{SNR}_{\text{peak}}$ a reasonably large $z_0$ is required, which can lead to axial frequency shifts. They are caused by electric anharmonicities and even-order magnetic field inhomogeneities in the center of the trapping potential due to the large $r_+$ at the end of the PnA method. Consequently, a highly harmonic electrostatic trapping potential and small magnetic field inhomogeneities are necessary.  

\begin{flushleft}
\textbf{Setup}
\end{flushleft}

\begin{figure*}
\includegraphics[width=0.7\textwidth]{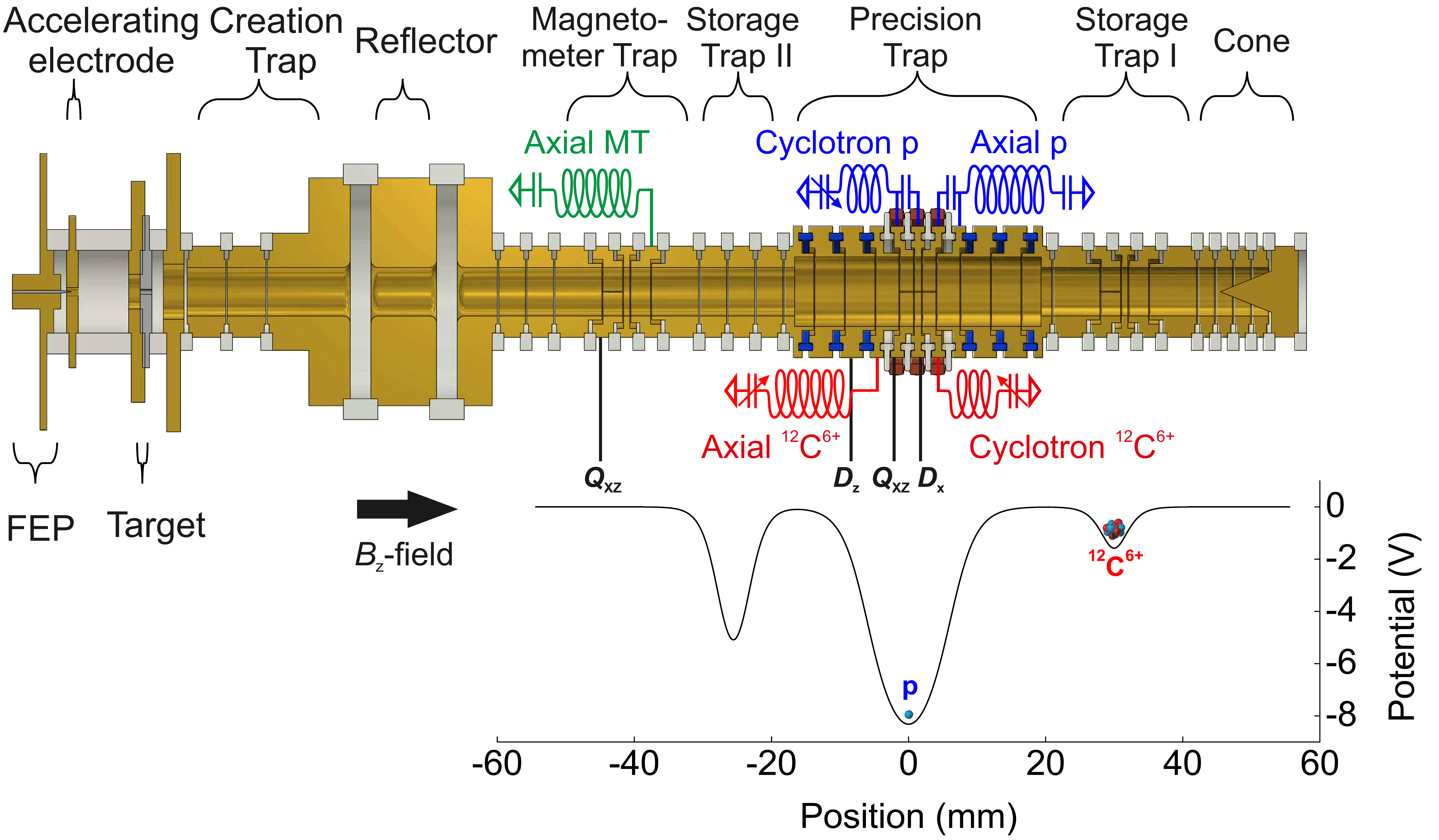}
\caption{\label{Traptower_large} Sketch of the complete trap tower including all detection systems and excitation lines as well as the resulting potential along the $z$-axis at the center of the electrodes of the $\text{LIONTRAP}$ experiment during the proton mass measurement campaign at the lower right. Besides the precision trap (PT) four other traps are shown: two storage traps (ST-I, ST-II), the magnetometer trap (MT), and the creation trap (CT). Additionally, there are several transport electrodes. The green tank circuit is the axial detection system for the magnetometer trap, whereas the two blue circuits are designed for the proton and the red ones are for the carbon ion. The black connections symbolize the four excitation lines. The cone, at the upper side of ST-I, is part of a cleaning technique to remove unwanted ions, which can be utilized in the future. In comparison to the proton mass paper~\cite{PhysRevLett.119.033001} two traps are renamed for clarification: the former measurement trap changed to precision trap and the reference trap is renamed to magnetometer trap.}
\end{figure*}

The $\text{LIONTRAP}$ experiment is the direct successor experiment of the former bound-electron \textit{g}-factor of highly charged ions experiment (\textit{g}-factor HCI) in Mainz~\cite{PhysRevLett.107.023002,PhysRevLett.110.033003,Nature2016,Nature2014,0953-4075-48-14-144032}. We developed a purpose-built Penning-trap stack as well as detection circuits optimized for CFR measurements of light ions. The superconducting magnet and the cryogenic reservoirs for liquid helium and nitrogen of the original experiment were reused. Our experimental approach requires single trapped ions and long storage times up to months. A very good vacuum is required to achieve this. To this end, the whole cylindrical Penning-trap tower is located in a hermetically sealed trap chamber. Cryopumping leads to a vacuum better than $10^{-17}\,\text{mbar}$ and storage times in excess of several months, which can be concluded by the lack of charge exchange with highly charged ions. The trap chamber itself is placed within the homogeneous region of the magnetic field of a $3.8\,\text{T}$ superconducting magnet. The Penning-trap tower consists of 38 cylindrical electrodes, see Fig.~\ref{Traptower_large}. Except of the PT, most electrodes are reused from the former experiment, the MT as well as the miniature electron beam ion source (mEBIS), including the creation trap (CT), the reflector, the electron gun and the target holder. 

\begin{flushleft}
\textbf{Detection system}
\end{flushleft}

In total, five different detection systems are connected to the electrodes as described in Fig.~\ref{Traptower_large}: one tank circuit for the detection of the axial frequency of the ion in the MT, and four tank circuits for the PT. Separate detection systems are attached to the respective electrodes of the PT for the proton's and the carbon ion's axial and modified cyclotron frequencies. The operation of four different tank circuits is necessary, because it is currently technically not possible to adjust high-$Q$ tank circuits over the large frequency range required. An overview of the properties of the five detection systems is given in Table~\ref{detectionsys}.

To prevent position shifts of the ions, which are hard to determine, the electrode voltages of the PT are set to the same value for both ions. Therefore, the axial frequency ratio of the carbon ion and proton is fixed to the charge-to-mass ratio of these two ions. To get both ions in resonance with their respective tank circuit, the carbon axial resonator in the PT is equipped with a voltage-variable capacitor that allows fine-tuning of its resonance frequency to the ion's axial frequency in a range of $5\,\text{kHz}$~\cite{Sven2012,10.1063/1.4967493}. The same applies for the modified cyclotron frequencies of the two ions, since the magnetic field of the superconducting magnet cannot easily be tuned to fit the ions' modified cyclotron frequencies to the resonator resonance frequency. In total, the resonance frequencies of the four individual tank circuits need to fit to the respective frequencies of the ions to determine the CFR of both ions at the same electric trapping potential and magnetic field.

The detection system of the magnetometer trap did not work properly and was not used during the first measurement campaign. Therefore, it was not possible to perform phase-sensitive measurements for both ion species simultaneously. However, it has meanwhile been repaired and can be used for the following mass measurement campaigns. 

Additionally, three excitation lines are connected to the PT. The quadrupole excitation $Q_{xz}$ is used for sideband rf-drives for cooling and to drive the double dips of the ions. Furthermore, it is required for the PnA method to transfer the modified cyclotron phase to the axial phase. The $Q_{xz}$ drive is connected to one half of the lower first correction electrode. It should be noted, that this drive is not a pure quadrupole excitation, but contains a significant dipole contribution.
The dipole excitation line $D_x$ is connected to one half of the ring electrode. It is used for the excitation in radial direction, which is required for the isolation of single ions. The axial dipole excitation $D_z$ is connected to the outer correction electrode. This excitation line is used for axial sweep excitations, which are also required to prepare single ions. Furthermore, the two dipole excitations ($D_x$ and $D_z$) can shape the $Q_{xz}$ towards a more precise quadrupole excitation by cancelling the unwanted dipole contributions. However, the two dipole excitations were shorted to ground during the measurement campaign to avoid excess noise entering via these lines. Another $Q_{xz}$ quadrupole drive is connected to the MT to enable the PnA method in this trap, which is sufficient since the required dipole excitation is carried by the dipole part of the $Q_{xz}$ excitation.

All our excitation lines are connected with GaAs transistor switches (SW-239, MACOM Technology Solutions~\cite{MACOM}) in the cryogenic electronic section to suppress residual rf-noise coupling to the trap electrodes~\cite{Sven2012}. If they are closed, their typical suppression at $700\,\text{kHz}$ is $50 \,\text{dB}$ and at $30\,\text{MHz}$ it is $20 \,\text{dB}$. When recording the double-dip for the modified cyclotron and magnetron frequency determination the switches are open. The systematic effects due to rf-noise at the dips are small and the required precision for these auxiliary measurements is in the order of $50\,\text{mHz}$. However, for the PnA measurement it is important to close the switches to avoid any additional noise during the high-precision determination of $\nu_+$. 

\begin{table*} 
\begin{center}
\caption{Summary of the characterisation of the five different detection systems. The inductance $L$, capacitance $C$, effective electronic electrode distance $D_{\text{eff}}$, the signal-to-noise ratio (SNR), the resonance frequency $\nu_{\text{res}}$ and the measured quality factor ($Q-$value) are listed. The SNR is the ratio of the maximum thermal noise of the resonator compared to the thermal background noise of the amplifier. The listed capacitances are only the ones from the resonators themselves. The total capacitances are larger since the resonators are coupled to the trap electrodes. The $Q$-value is calculated by $Q = \nu_{\text{res}} / \Delta \nu$, where  $\Delta \nu$ is the FWHM of the thermal noise signal of the resonator on the frequency spectrum. The parallel ohmic resistance of the resonators $R_{\text{p}} = 2 \pi  Q  L \nu_{\text{res}}$ and their corresponding cooling time~$\tau$ at $\nu_z=\nu_{\text{res}}$ with $\tau(\nu_{\text{res}}) = m D^2_{\text{eff}} / \left( q^2  R_{\text{p}} \right)$ are listed, too. The energy damping of the corresponding motion is proportional to $\exp\left(-t/\tau\right)$.}\label{detectionsys}
\begin{ruledtabular}
\begin{tabular}{c c c c c c}
Detection  system  & Axial MT & Axial p & Axial $^{12}$C$^{6+}$\footnote{\label{varactor}These resonators are connected to a voltage-variable capacitor to fine-tune their exact resonance frequency. With the frequency also the quality factor and thus the cooling time constant varies.} & Cyclotron p$^{\text{\ref{varactor}}}$ & Cyclotron $^{12}$C$^{6+\text{\ref{varactor}}}$ \\
 \hline
$L\,\text{(mH)}$ & 2 & 1.65 & 3.36 &  $5.6 \times 10^{-4}$ & $2.4\times 10^{-3}$ \\
$C\,\text{(pF)}$ &  15 &  9 & 11 & 2.4 & 2 \\
$D_{\text{eff}}$ (mm) & 7.4 &  11.6 & 14.0 &  16.3 & 28.7 \\
SNR (dB) & 24 &  13 &  14 & 4 & 4 \\
$\nu_{\text{res}}$  $\left(\text{kHz}\right)$ & 671 & 739 &  524 $-$ 528 &  57\,200 $-$ 57\,500 & 28\,400 $-$ 28\,900 \\
$Q$-value & 20\,300 &  4450 & 2200 $-$ 2900   & 190 $-$ 520 & 200 $-$ 610 \\
$R_{\text{p}}$ (M$\Omega$) &   172 & 34 & 25 $-$ 31  & 0.04 $-$ 0.10 & 0.09 $-$ 0.27 \\  
$\tau \left(\text{s}\right)$  &  $6\times 10^{-3}$\footnote{The cooling time is calculated for $^{12}$C$^{6+}$.} &  0.25 &   0.17 $-$ 0.13 &  450 $-$ 170  &   193 $-$ 66  \\
\end{tabular}
\end{ruledtabular}
\end{center}
\end{table*}

\begin{flushleft}
\textbf{Creating and storing single ions}
\end{flushleft}

The carbon ion as well as the proton are created in the mini-EBIS, which is located at the lower side of the trap chamber~\cite{1742-6596-58-1-021,0953-4075-43-7-074016}. Oscillating electrons ablate atoms from a carbon nanotube-filled PEEK target (TECAPEEK)~\cite{Tecapeek} with a $700\,\mu\text{m}$ hole in the center for the electron beam. The carbon nanotubes are necessary to guarantee electrical conductivity. At a voltage difference of about $700\,\text{V}$ between the accelerating electrode and the field emission point (FEP), the FEP starts emitting electrons with a current up to a few hundred nanoamperes. Protons are created at a beam energy of $-90\,\text{V}$ applied for $4\,\text{s}$ at the FEP, whereas for carbon ions a voltage of $-900~\text{V}$ for $4\,\text{s}$ is applied. This meets the criterion that the largest cross section for the production of $^{12}\text{C}^{6+}$ is achieved at a beam energy being a factor of 2.5 larger than the ionization energy~\cite{Valyi1977}, see Table~\ref{binding_c}. The reflector electrode is set on voltages of $-100\,\text{V}$ and $-1000\,\text{V}$ for the proton and the carbon ion production, respectively. During a creation process for carbon ions, all charged states of carbon are produced and also ions with lower ionization energies, such as protons. Basically only protons and H$_2^+$ molecules are produced during the creation process for protons. All ions are stored in the creation trap. Then the whole ion cloud is adiabatically transported to the PT. On average, in the order of ten $^{12}\text{C}^{6+}$ or hundreds of protons are detected in the PT after one respective creation cycle.

For removing all unwanted ions the so-called ``magnetron cleaning" is used. In this process broadband white noise from $0-10\,\text{kHz}$ is applied via the $D_x$ excitation line. At the same time only the magnetron motion of the ion of interest is cooled via sideband coupling to the axial tank circuit at the $q/m$-sensitive frequency $\nu_{\text{rf}-} = \nu_- + \nu_z$. Consequently, the magnetron motion of all other ions increases until they hit the surfaces of the electrodes and get lost. The whole process lasts for about $15\,\text{min}$ to make sure that no unwanted ions remain in the trap.
An additional method to prepare ions of a single species is to apply a broadband axial excitation. This excitation is in the range well above the magnetron frequencies and below $2 \cdot \nu_z$, typically from $100\,\text{kHz}-1\,\text{MHz}$, except the range of $\pm 25\,\text{kHz}$ around the axial frequency of the particle of interest. After this excitation the trap potential is dipped towards $U_{\text{R}} \approx -100\,\text{mV}$ for one second. The ions which had been previously axially excited are removed by this procedure. 

If several ions of the same type remain, their modified cyclotron motion is excited and their individual signals can be observed as peaks on the corresponding cyclotron resonator. After that the trapping potential is lowered step-wise, while observing the single peaks of the individual ions in the frequency spectrum of the resonator. The modified cyclotron frequencies of simultaneously trapped ions of the same type with different kinetic energies deviate slightly due to special relativity and residual magnetic field inhomogeneities. If the trapping potential is low enough, some ions are not confined anymore and escape. This is observed by the disappearance of their signal in the spectrum. After that the trap is set to the original potential and all three motions of the remaining ion or ions are cooled. This procedure is repeated until one single ion remains. It is possible to determine the number of thermalized ions of the same species $N$ via the FWHM of the dip signal at the axial detection systems. The FWHM in the power spectrum scales for small numbers $N$ linearly with the number of thermalized protons and carbon ions, respectively~\cite{doi:10.1063/1.360947}. This relation only holds for a common-mode motion of all $N$ ions, resulting in a FWHM given by:
\begin{equation}
\Delta \nu_z = \frac{N}{2 \pi}\frac{1}{\tau_z} = \frac{N}{2 \pi} \frac{R_{\text{p}} q^2}{m D_{\text{eff}}^2} \quad ,
\end{equation}
where $\tau_z$ is the cooling time constant of a single ion on the corresponding axial resonator.

The single ion is adiabatically transported to the $\text{ST-I}$. Subsequently, the whole process is repeated to create another single ion in the PT. The creation of two single ions in two different traps can potentially lead to captured electrons between the two traps, which can lead to a distorted electrical potential of the traps. To eliminate these electrons, the potential of the PT is lowered in a way, that the electrons are transported through it and stored at the lower side of the PT. After that the electron cloud is transported to the CT and dumped to the wall.

\section{The doubly compensated precision trap} \label{chap_MT}

\begin{flushleft}
\textbf{Realization of a highly-harmonic seven-electrode cylindrical Penning trap}
\end{flushleft}

\begin{figure}
\includegraphics[width=0.3\textwidth]{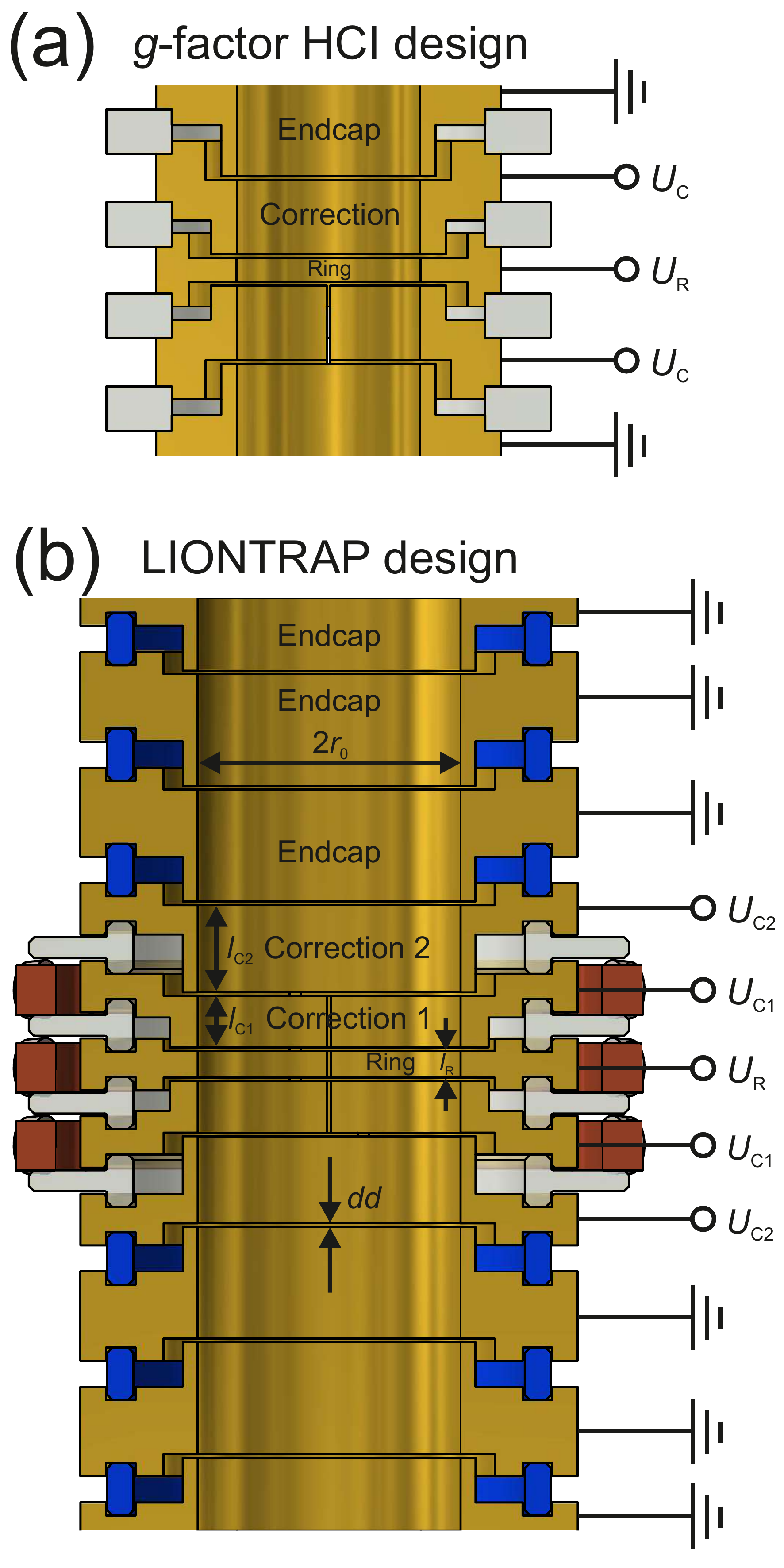}
\caption{\label{Shrinking} Picture (a) shows a vertical cut through the cylindrical electrodes (gold) and the quartz rings (gray) of a typical five electrode cylindrical Penning trap, here, of the former \textit{g}-factor HCI precision trap. The configuration can cause a possible radial misalignment due to the larger thermal contraction of the trap electrodes in comparison with the quartz spacers. Such a misalignment is avoided with the seven electrode design of the precision trap, shown in the lower illustration (b) including all adjustment parameters. There the electrodes shrink onto the sapphire rings (blue) due to a larger thermal expansion coefficient, resulting in a self-alignment of the electrodes. The split electrodes are arranged on T-shaped quartz glass rings. Quartz glass was used instead of sapphire due to the easier manufacturing. The copper rings (brown) are used for the fixation of the three split inner electrodes.}
\end{figure}

Cylindrical Penning traps have five-electrode designs~\cite{GABRIELSE19841,10.1063/1.102084,GABRIELSE1989319,Schneider1081,Smorra2017}. They consist of one ring electrode, one pair of correction electrodes and one pair of endcap electrodes to shape a harmonic potential and cancel leading order anharmonicities in the electric potential. Several approaches for even higher-harmonic traps have been discussed in literature~\cite{FEI1999431,SIKDAR2013174}. For the PT at LIONTRAP such an even more harmonic Penning trap has been designed and constructed~\cite{Florian2015}. Here, a second pair of correction electrodes is added, resulting in a seven-electrode cylindrical Penning trap. The two endcap electrodes are segmented to guarantee a reliable adiabatic transport of the ions, see Fig.~\ref{Shrinking}. We consider our trap as the first doubly compensated seven electrode trap, not counting the so-called ``Preparation Trap" of the ISOLTRAP experiment~\cite{Mukherjee2008}. 

In five-electrode cylindrical Penning traps the coefficient $C_2$ (see Eq.~(\ref{pot_tay})), which characterizes the strength of the trapping potential, can be split into two contributions:
\begin{equation}
C_2 = D_2 \cdot \text{TR} + E_2 \quad .
\end{equation}
$U_{\text{C}}$ is the potential, which is applied to both correction electrodes, and $\text{TR}=U_{\text{C}}/U_{\text{R}}$ is the so-called tuning ratio. Since $\nu_z$ is proportional to $\sqrt{C_2}$, it is favorable that $C_2$ is independent of the correction voltage, resulting in $D_2 =0$. Such a trap is called an orthogonal trap. For two correction electrode pairs a double orthogonality $D_{2,1}=D_{2,2}=0$ would be favorable:
\begin{equation}
C_2=D_{2,1} \frac{U_{\text{C}1}}{U_{\text{R}}}+D_{2,2} \frac{U_{\text{C}2}}{U_{\text{R}}} + E_2 \quad .
\end{equation}

However, this double orthogonality cannot be achieved. Therefore, we aim for a slightly weaker condition in our trap, which we call combined orthogonality:
\begin{equation}
D_2^{\text{comb}} \equiv D_{2,1} \frac{U_{\text{C}1}}{U_{\text{R}}}+D_{2,2} \frac{U_{\text{C}2}}{U_{\text{R}}} = 0 \quad .
\end{equation}

In this way, the axial frequency stays constant, when $U_{\text{C}1}$ and $U_{\text{C}2}$ are scaled by the same factor. There are typically three degrees of freedom in a five-electrode cylindrical Penning trap: the compensation voltage $U_{\text{C}}$ as well as the lengths of the ring and the correction electrodes. They are optimized to reach an orthogonal and compensated trap with $C_4=C_6=D_2=0$. In our trap design the additional pair of correction electrodes provides two more degrees of freedom: their lengths and the applied voltages, see Table~\ref{degfree}. Therefore, it is possible to design a doubly compensated trap with
\begin{equation}
C_4=C_6=C_8=C_{10}=D_2^{\text{comb}}=0 \quad.
\end{equation}
The radius is fixed to $r_0=5~\text{mm}$, which is a balance between a reasonably high SNR$_{\text{peak}}$, since $D_{\text{eff}}$ scales linearly with the trap radius, and a small image charge shift, which scales with $1/r^3$, a systematic shift described further in section~\ref{mp_stat}. Furthermore, a larger radius diminishes the effects of misalignment, deformation and machining imperfections. Additionally, unwanted effects caused by non-uniform work functions due to varying crystal orientations as well as non-conductive islands on the surface of the gold-plated electrodes that can charge up during the ion creation process are reduced.

These non-conductive islands are so-called patch potentials and lead to a distorted electric field. In our trap patch potentials are $U_{\text{patch}} < 10\,\text{mV}$ for the PT and thus more than an order of magnitude smaller than in the precision trap of the former \textit{g}-factor HCI experiment, for details see Appendix~\ref{patch_app}.

For the gap between the electrodes $dd=0.14~\text{mm}$ is chosen. This is a trade-off between the capacitances, which limit the parallel resistance of the tank circuit, and electric field imperfections.
\begin{table} 
\begin{center}
\caption{Degrees of freedom for the precision trap. An overview of fixed and optimized trap parameters is given. $U_{\text{R}}$ has a design value between $- 14\,\text{V}~\text{and}~0\,\text{V} $ to provide axial frequencies of several hundreds of kHz, which enables the use of the ultra-stable voltage supply UM1-14~\cite{birgit2011,Sven2012}. The design values for the dimensions represent the final lengths at $4\,\text{K}$, including the different layers of material. In good approximation, the length of the endcap electrodes is assumed to be infinitely long.}\label{degfree}
\begin{ruledtabular}
\begin{tabular}{l c c}
& Trap parameters & Design values \\
 \hline
\multirow{3}{*}{ fixed }  & $r_0$ & 5.000~mm \\
& distances between & \multirow{2}{*}{ $0.140\,\text{mm}$} \\
&  electrodes  $dd$  & \\  
\hline
\multirow{6}{*}{\shortstack{optimized by \\ simulations~\cite{Florian2015}} }  &  length of ring $l_{\text{R}}$ & $1.047\,\text{mm}$ \\
&  length of corr. el. 1 $l_{\text{C1}}$ & $2.000\,\text{mm}$ \\
 &  length of corr. el. 2 $l_{\text{C2}}$ & $3.355\,\text{mm}$\\
 & voltage of corr. el. 1 $U_{\text{C1}}$ & $0.963\,57\,U_{\text{R}}$\\
& voltage of corr. el. 2 $U_{\text{C2}}$ & $0.815\,55\,U_{\text{R}}$\\
\end{tabular}
\end{ruledtabular}
\end{center}
\end{table}

Besides the larger radius and the different electrode lengths also the electrode design changed to minimize their capacitance, which is about $10\,\text{pF}$ for the PT electrodes. Due to the mismatch of the integrated thermal expansion coefficients of the quartz rings ($\eta^{4\,\text{K} - 300\,\text{K}} _{\text{quartz}} \approx \eta^{4\,\text{K} - 300\,\text{K}}_{\text{sapphire}} = 1.0008$) and the copper electrodes ($\eta^{4\,\text{K} - 300\,\text{K}}_{\text{Cu}}~=~1.0032$), the mechanical design was modified compared to the previous trap designs. Now, the T-shaped quartz rings are contained inside the copper electrodes, so that these self-align when shrinking onto the quartz rings, see Fig.~\ref{Shrinking}. Furthermore, the distance between the vertically split electrodes is also $0.140\,\text{mm}$ to guarantee a virtually closed surface.

The electrodes consist of OFHC-copper (oxygen free high-conductivity copper) with a $2~\mu$m silver diffusion barrier in between and  a $11 \,\mu \text{m}$ gold layer. OFHC-copper is chosen to prevent severe magnetic field disturbances due to paramagnetic oxygen contaminations at $4\,\text{K}$~\cite{VanDyck2001}.

In the future, a material with lower electric conductivity could be chosen to reduce the eddy current lifetime. The electrodes have been manufactured in the workshop of the Institute of Physics at the Johannes Gutenberg-University Mainz. The galvanic silver and gold plating was done by the company Drollinger~\cite{drollinger}.

The total geometric uncertainty of the electric potential is $\pm 20 \,\mu$m. This uncertainty includes the machining tolerances of the electrodes, the possible deformation of the split electrodes, the uncertainty of the gold and silver layers as well as the alignment uncertainty of the mounted electrodes, unequal work functions of the material and patch potentials. The height of the quartz rings has an uncertainty of $1~\mu$m.

\begin{figure}
\includegraphics[width=0.48\textwidth]{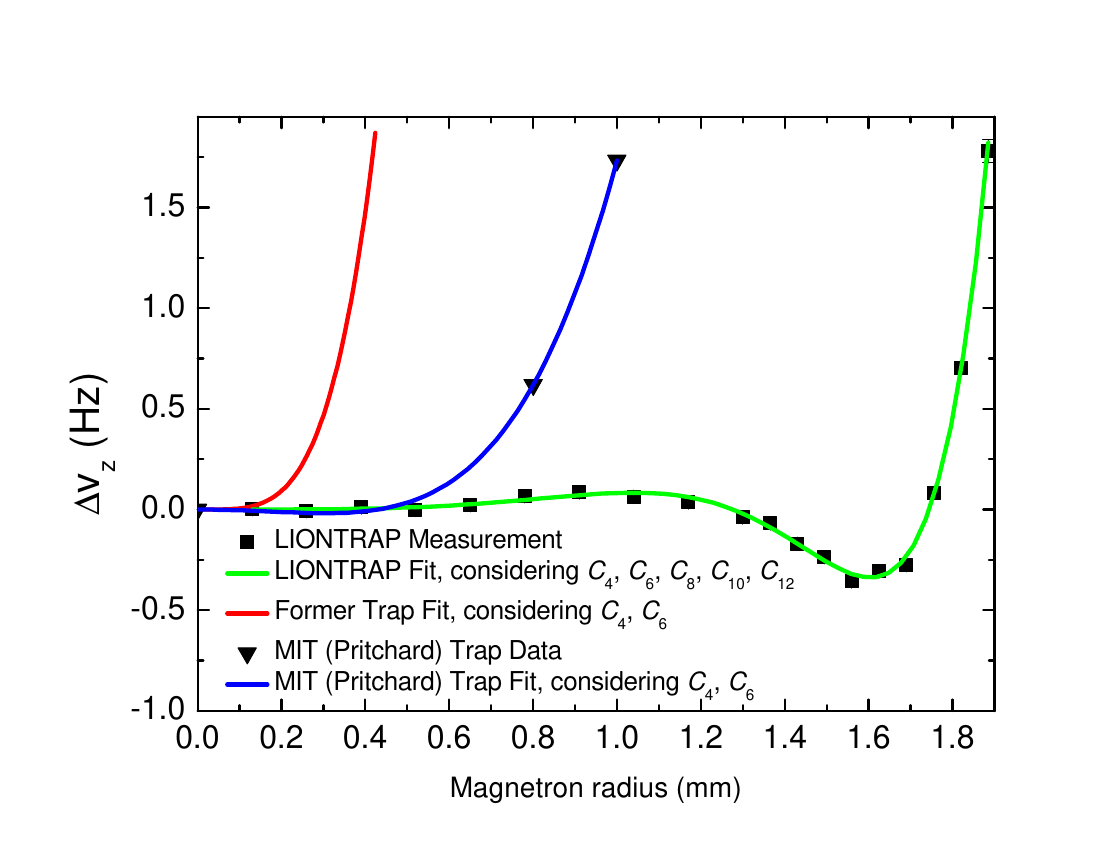}
\caption{\label{opt_Tr} Observed axial frequency shift $(\Delta \nu_z=\nu_z^{\text{exc}}-\nu_z^{\text{thermal}})$ following a magnetron excitation at $\nu_z \approx 525\,\text{kHz}$. Our trap is significantly more harmonic compared to our former \textit{g}-factor HCI trap~\cite{0953-4075-48-14-144032}. It is even more harmonic than the most harmonic hyperbolic Penning trap from MIT/FSU~\cite{Rainville2003}. The MIT/FSU Penning trap has the following trap parameters: $\nu_z^{\text{MIT}} \approx 212\,\text{kHz}$, $d_{\text{char}}^{\text{MIT}} = 5.49 \,\text{mm}$ and $C_2=1$. To achieve a reasonable comparison with the MIT data, we scaled their axial frequency to our value.}
\end{figure}

The geometric uncertainties limit the harmonicity of the trap. This harmonicity can be optimized in-situ by applying proper voltages to the two pairs of correction electrodes. Systematic studies of this optimization are performed to achieve the required highly harmonic electrical trapping potential. Moreover, the performance of the trap predicted by the simulation can be checked. This is done by determining the residual electrostatic anharmonicities, expressed with the even $C_i$-coefficients ($i \geq 4$), see Eq.~(\ref{pot_tay}) via the shift of the axial frequency due to an excitation of the magnetron motion. The simulated slopes and the measured ones are in remarkable agreement, for details see Appendix~\ref{harmonicty_trap}. Table~\ref{harmon_c} summarizes the coefficients in the optimal configuration. 
\begin{table} 
\begin{center}
\caption{Comparison of the even order coefficients $C_n$ for the in-situ optimized doubly compensated presicion trap (PT) of $\text{LIONTRAP}$ and the singly compensated precision trap (PT) of the former \textit{g}-factor HCI experiment. Due to the large axial shifts it was not possible to determine the higher order coefficients for the former trap.}\label{harmon_c}
\begin{ruledtabular}
\begin{tabular}{c c c}
Coefficent &  PT ($\text{LIONTRAP}$) & PT (\textit{g}-factor HCI)\\
 \hline
 $C_2$ & -0.5997 & -0.5504 \\
$C_4$ & $0.07(1.29) \times 10^{-6} $	&   $ 0(1) \times 10^{-5} $	\\ 
$C_6$ & $	-4.3(4.6) \times 10^{-5} $ &  $1.6(1) \times 10^{-2} $	\\
$C_8$ & $ 9.8 (65.0) \times 10^{-5}$ & \rule{1cm}{0.4pt} \\
$C_{10} $ &  0.0115(42)  & \rule{1cm}{0.4pt} \\
$C_{12} $ &  0.062(10)  & \rule{1cm}{0.4pt} 
\end{tabular}
\end{ruledtabular}
\end{center}
\end{table}
To find the optimum voltage configuration with $\left(p_1=p_2=0\right)$ we perform a two dimensional scan by varying $U_{\text{C1}}$ and $U_{\text{C2}}$ independently. After optimization, the remaining uncertainty of $C_6$ is a factor of 4000 smaller in the PT of $\text{LIONTRAP}$ than in the former trap of the \textit{g}-factor HCI experiment, see Fig.~\ref{opt_Tr}. Additionally, all coefficients up to $C_8$ are zero within the error bars. 

\begin{figure}
\includegraphics[width=0.48\textwidth]{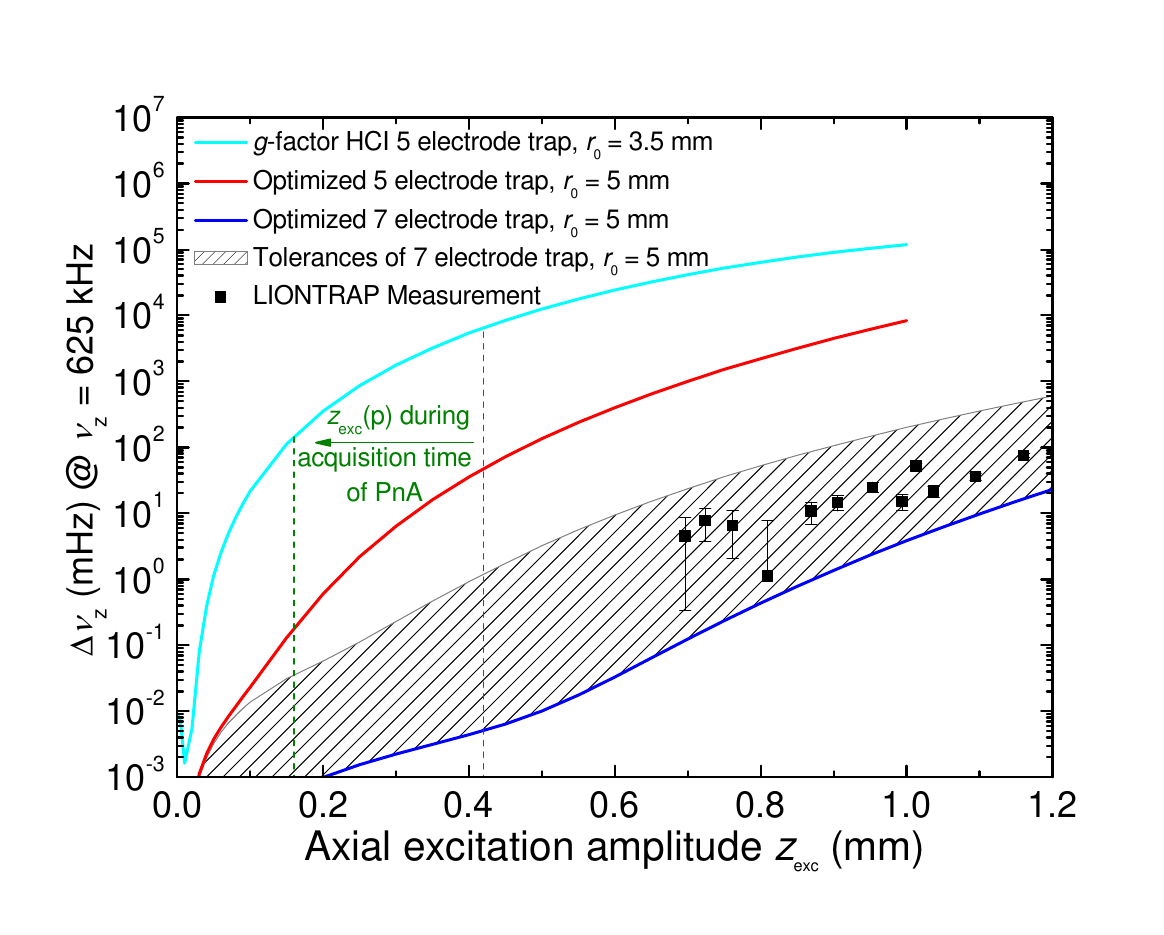}
\caption{\label{axial_sim} Comparison of calculations and measurements of the axial shift in dependence of the axial amplitude. The significantly larger shift of the former \textit{g}-factor HCI trap in comparison to the optimized five electrode trap is caused by a flaw in the trap calculation of the former \textit{g}-factor HCI trap. The tolerances arise from allowing variations of the lengths of the electrodes, distances and diameter by $20\,\mu\text{m}$. A shift for smaller amplitude is hard to measure due to the systematic uncertainty of the dip measurement. The data points, which are determined from different excitation and delay times, are in good accordance with the simulations. The improvement between $\text{LIONTRAP}$ and the former trap is more than three orders of magnitude. The two vertical green dashed lines show the excitation amplitude of the proton after the second PnA pulse during the acquisition time of the axial peak. These parameters are used during the measurement campaign for the determination of the proton's atomic mass.}
\end{figure}

Another cross check for the electrostatic harmonicity of the trap is to apply an excitation at the ion's axial frequency and determine the shift of this frequency in dependence of the excitation energy. The frequency is determined via the axial phase of the excited ion signal on the resonator. This phase measurement has been performed with an axial frequency slightly above the resonator frequency to increase the cooling time constant and thus the phase evolution time. Different excitation lengths and different phase evolution times are chosen for the measurement. This axial frequency shift is compared to simulations, see Fig.~\ref{axial_sim}. Our trap shows a three orders of magnitude smaller axial shift at $z_{\text{exc}}~=~1 \,\text{  mm}$ in comparison to the former \textit{g}-factor HCI trap design. The observed shift is within the manufacturing tolerances.

With this voltage setting the trap is sufficiently harmonic that it does not restrict the precision of our measurements and relative systematic shifts of the cyclotron frequency are below $10^{-13}$. Moreover, it enables us to perform phase-sensitive measurements of the modified cyclotron frequency with a single proton for the first time, which so far have been limited by axial frequency shifts due to trapping anharmonicities. Here, we can excite the axial motion of the ion at the end of the PnA cycle to large enough amplitudes for achieving a sufficient peak signal. A peak SNR of approximately $14\,\text{dB}$ is necessary to achieve a technical readout jitter below $15^\circ$~\cite{Florian2015}, which is relevant for the PnA method. This corresponds to an averaged axial amplitude of $ z_{\text{exc}} = 260\,\mu\text{m}$ for the proton during the acquisition time of the axial peak. A higher SNR for the proton was not possible due to the quadratic magnetic field component. This inhomogeneity together with the increased modified cyclotron radius after the second PnA pulse led to an additional axial frequency shift. For a larger axial excitation amplitude this shift would additionally increase the phase jitter.

\begin{flushleft}
\textbf{Magnetic field}
\end{flushleft}

\begin{figure}
\includegraphics[width=0.48\textwidth]{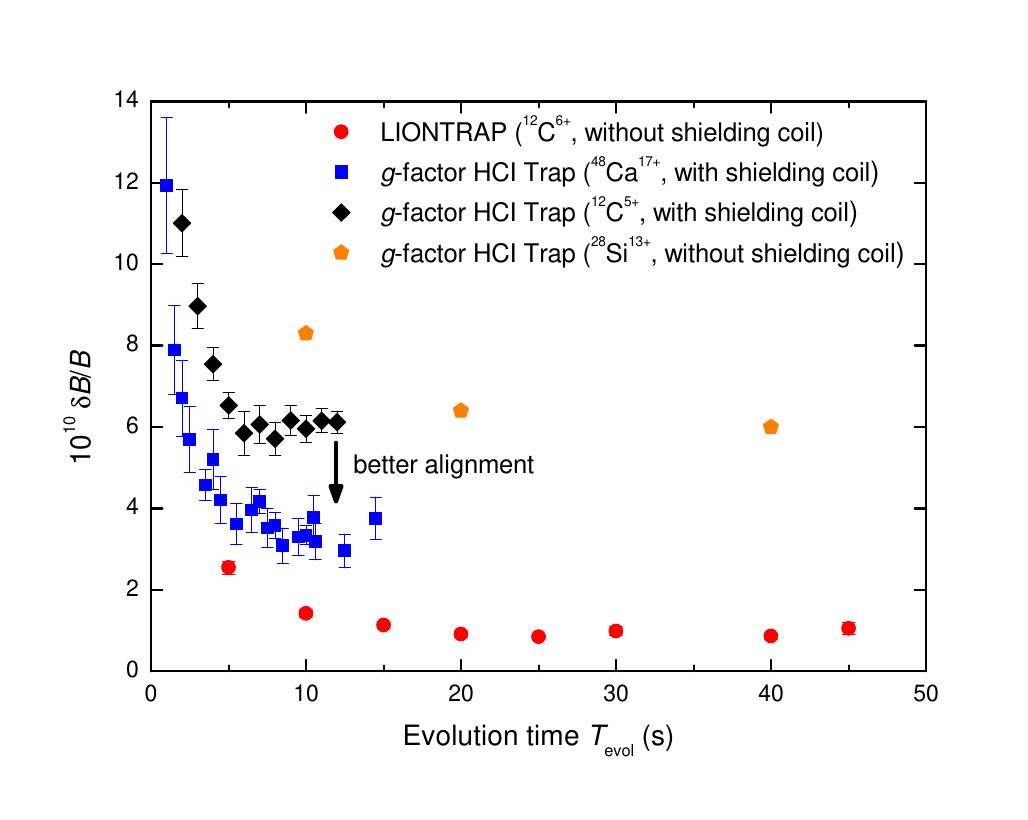}  
\caption{\label{mag_stab} Comparison of the magnetic field stabilities of the former \textit{g}-factor HCI experiment, determined using a $^{28}$Si$^{13+}$, $^{48}$Ca$^{17+}$, and $^{12}$C$^{5+}$ ion, and of the $\text{LIONTRAP}$ experiment using a $^{12}$C$^{6+}$ ion. During the $^{48}$Ca$^{17+}$ ion and the $^{12}$C$^{5+}$ ion measurements a superconducting closed self-shielding compensation coil was installed. Longer evolution times are not possible due to unwrapping errors. The total cycle time in between two successive determinations of $\phi_+^i$ is given by $T_{\text{cycle}} = T_{\text{evol}} + T_{\text{cool}}$, where  $T_{\text{cool}} \approx 45\,\text{s}$ is the combined cooling time for the axial and the modified cyclotron mode of the ion after the phase determination.}
\end{figure}

The magnetic field stability is the dominant source of statistical fluctuations of the mass measurement. It can be measured via:
\begin{equation}
\frac{\delta B}{B} = \frac{\delta \nu_c}{\nu_c} \approx \frac{\delta \nu_+}{\nu_+}=\frac{\delta \phi_+}{\phi_+}= \frac{\delta \phi_+}{360^\circ \cdot \nu_+ \cdot T_{\text{evol}}} \quad , \label{magistabi}
\end{equation}
where $\delta \phi_+$ is the variation of the total modified cyclotron phase in degree between successive phase measurements, given by $\delta \phi_+ \equiv \text{std}(\phi_+ )= \text{std(diff}(\phi_+^i , \phi_+^{i+1} ))/\sqrt{2}$, which is the differential change in the phase of $\nu_+$ in subsequent measurements $i$ and $i+1$, and $T_{\text{evol}}$ is the evolution time of the measurement. For the proton mass campaign we decided for $ T^{\text{max}}_{\text{evol}} = 10\,\text{s}$. Different evolution times are chosen to extract the magnetic field stability, see Fig.~\ref{mag_stab}. The stability is determined by repeating several PnA cycles with identical $T_{\text{evol}}$ and recording $\phi_+^i$. 

The trap of the former \textit{g}-factor HCI experiment was initially shielded by the  	built-in self-shielding coil of the magnet. The magnetic field fluctuations for $T_{\text{evol}}=20\,\text{s}$ were determined with $^{28}\text{Si}^{13+}$ to be $\delta B / B \approx 6 \times 10^{-10}$~\cite{Sven2012}. Later, a home-made closed superconducting self-shielding compensation coil was placed directly around the trap chamber to reduce the magnetic field fluctuations at the trap center~\cite{Anke2013,Florian2015}. Even though the coil showed a large shielding factor against external field fluctuations, the stability was initially improved at $T_{\text{evol}}=10\,\text{s}$ by only a small factor, and for $T_{\text{evol}}=20\,\text{s}$ the stability was basically unchanged. Later, the stability was improved by a factor of two due to the better alignment of this coil~\cite{Anke2013,Florian2015}. This coil was removed in the $\text{LIONTRAP}$ setup to implement $B_1$ and $B_2$ shim coils, see Appendix~\ref{mag_shim_coils}. Surprisingly, the stability improved once more by a factor of three compared to the previous setup and is now $ \delta B / B \approx 1 \times 10^{-10} $. The reason for this improvement is unclear. Possible reasons might be the removed ferromagnetic nickel electrode from the former \textit{g}-factor HCI experiment or the elimination of a resonator with a superconducting housing, which could lead to magnetic field instabilities due to micro-vibrations.

The quadratic magnetic field inhomogeneity $B_2$ is the dominant magnetic inhomogeneity, which is responsible for energy dependent shifts in the three eigenfrequencies. The determination of $B_2$ is possible via an excitation of the modified cyclotron mode and the subsequent measurement of the shifts in the axial and modified cyclotron frequency. We can determine $B_2=-0.270(15)\,\mu\text{T/mm$^2$}$. An asymmetric trapping potential is another way to determine the  leading  magnetic inhomogeneities. They are determined to $B_2=-0.286(12)\,\mu\text{T/mm$^2$}$ and  $B_1=0.925(15)\,\mu\text{T/mm}$. The result for $B_2$ is in excellent agreement with the one mentioned before. The details of these measurements can be found in Appendix~\ref{mag_app}. For the further analysis $B_2=-0.270(15)\,\mu\text{T/mm$^2$}$ is used.

\begin{flushleft}
\textbf{Ion temperature}
\end{flushleft}

The modified cyclotron temperature of the ion is a crucial parameter for CFR measurements, because it leads to a relativistic mass increase related to an additional systematic shift. This relative shift $\Delta \nu_+ / \nu_+ \approx -E_+/(m c^2)$ is especially large for the proton, as for a sideband thermalized ion it scales with $q/m^2$, when keeping the axial frequency constant. Therefore, the axial temperature $T_z$ of the ions should be low. To lower the axial temperature it is possible to apply negative feedback on the axial resonators~\cite{UrsoFBCool}. To this end, the phase of the signal of the detection system is shifted and fed back to the tank circuit.

We determined the temperature of the ions via two different methods: the lineshape of the dip at a large anharmonic electric quadrupole potential and via the jitter of the axial frequency after modified cyclotron excitation, for detail see Appendix~\ref{ion_temp_app}. The axial temperatures for the proton and the carbon ion are listed in Table~\ref{temp_values}. 

\begin{table} 
\begin{center}
\caption{Overview of the temperature of the axial motions for the proton and the carbon ion. ``Fb" represents negative electronic feedback applied and ``no Fb" stands for no electronic feedback applied. The difference in the case of the carbon axial temperature with applied feedback is due to a flaw in the temperature determination during the proton mass campaign, which is described in detail in Appendix~\ref{ion_temp_app}. Note, the proton and $^{12}\text{C}^{6+}$ have different axial frequencies, they are cooled via different tank circuits and their signal is amplified by different amplifiers. Therefore, the axial temperatures can be different due to frequency depended noise, different coupling of noise to the lines connected to their respective amplifier and due to different amplifier performances.}\label{temp_values}
\begin{ruledtabular}
\begin{tabular}{c c c c c}
$T_z$  (K) & \multicolumn {2}{c}{Proton}  & \multicolumn{2}{c}{Carbon}\\
 & no Fb & Fb & no Fb  & Fb   \\
 \hline
Reanalysis & 3.4(1.0) & 1.5(1.0) & 6.4(1.0)  & 4.5(1.4) \\
$m_{\text{p}}$ paper~\cite{PhysRevLett.119.033001}  & 4.2(1.0) & 1.7(1.0)  & 4.2(2.0) & 1.7(1.0)
\end{tabular}
\end{ruledtabular}
\end{center}
\end{table}

\section{Determination of $m_{\text{p}}$ } \label{mp_stat}

\begin{flushleft}
\textbf{Measurement cycle}
\end{flushleft}

The preferably simultaneous determination of the CFR of two ions is realized in the following way. After the determination of the cyclotron frequency of one ion in the PT, this ion is transported in the ST and the modified cyclotron frequency of the other ion, which was parked in the other ST and has been transported in the PT, is subsequently measured in the PT at the same position~\cite{ulmer15}. This shortens the switching time to one minute, while an interaction between the two ions is strongly suppressed. The time span between these two measurements should be as small as possible to reduce the magnetic field changes, which currently limit our statistical uncertainty. Therefore, the time between the modified cyclotron frequency measurements is optimized to be smaller than $5\,\text{min}$. Furthermore, the creation process of a single ion by an ion cloud from the mEBIS can lead to altered patch potentials on the electrode surfaces, which are completely avoided with our measurement method discussed here. These altered patch potentials would change the ion positions, which results in a systematically different magnetic field.

The proton and the carbon ion are neither a mass doublet nor a charge-to-mass doublet. In our setup the free cyclotron frequencies of the proton and the carbon ion are $\nu_c \left(\text{p}\right)\approx 58\,\text{MHz}$ and $\nu_c \left(^{12}\text{C}^{6+}\right) \approx 29\,\text{MHz}$, respectively. Using only one detection system requires applying a factor of two different ring voltages. Due to the inherent patch potentials on the electrode surfaces this can lead to different equilibrium positions of the ions within the trap, which causes a systematic effect on the CFR due to magnetic inhomogeneities. As an example, if the ring voltage is changed from $U_r \approx -9.8\,\text{V}$ to $U_r \approx -4.9\,\text{V}$, the electrostatic trap center is shifted by $\Delta L = 220\,\text{nm}$ along the axial direction for the case of one patch potential of $1\,\text{mV}$ located at the lower correction electrode 1. This shift already results in a systematic effect on the CFR of $ \Delta \text{CFR} / \text{CFR}=  \Delta B / B = \Delta L \cdot B_1 / B_0 \approx 6 \times 10^{-11}$, with the current magnetic field gradient $B_1$. Therefore, we use for the first time two different detection systems for the two axial frequencies. The CFR is measured at identical trapping potentials for both ions to guarantee identical equilibrium positions and thus identical magnetic fields.

\begin{figure}
\includegraphics[width=0.3\textwidth]{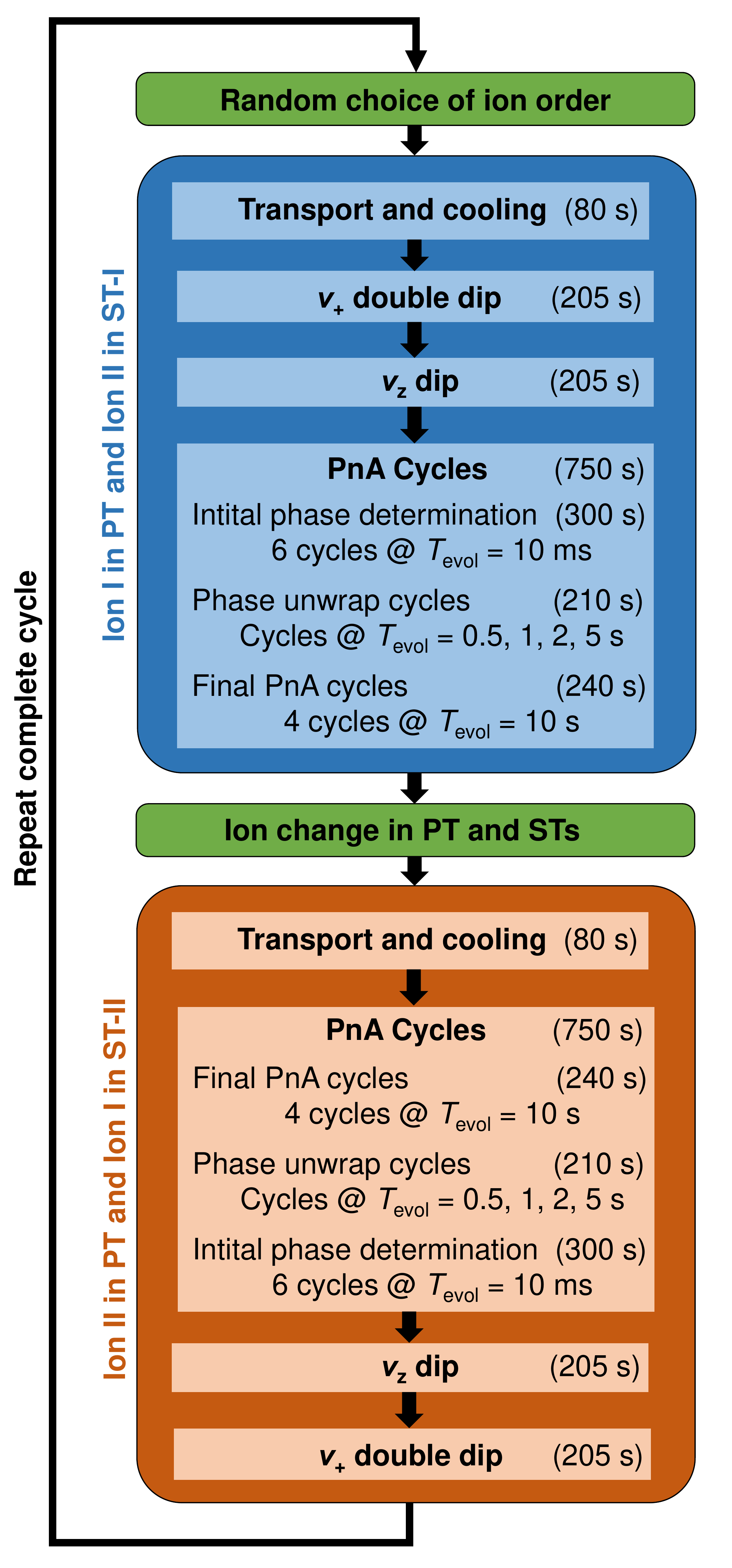}
\caption{\label{zeitverlauf} Sketch of the measurement cycle. After one cycle with the duration of $45\,\text{min}$ we get a relative statistical uncertainty for the proton's atomic mass of $1.8~\times~10^{-10}$. This is only a factor of two less precise than the CODATA 2014 literature value. The measurement itself is accomplished fully automatized over 24 hours a day. It is only interrupted by the helium and nitrogen filling of the reservoirs for the magnet and the apparatus. The total real measurement time amounts to approximately 300 hours.}
\end{figure}

The complete measurement cycle is shown in Fig.~\ref{zeitverlauf}. The ion to be measured first is chosen randomly in each cycle to exclude systematic errors such as a linear drift of the magnetic field for example due to heating effects caused by excitations during the PnA method~\cite{PhysRevLett.98.053003}. At the beginning, the first ion is transported into the PT and all three eigenmotions are cooled. The other ion is transported to one of the neighboring storage traps (ST). For both $\nu_c$ measurements the applied voltages in all three traps are set to identical values to guarantee truly equal electrostatic potentials. 

During the $\nu_c$ measurement in the PT, $\nu_+$ is initially measured via the double-dip technique, followed by the determination of $\nu_z$ via the dip technique. Finally, the modified cyclotron frequency is measured phase-sensitively. For the initial phase determination six PnA cycles are applied with $T_{\text{evol}}=10\,\text{ms}$, followed by cycles at evolution times of $0.5,~1,~2 \text{ and } 5\,\text{s}$ to allow proper phase unwrapping~\cite{PhysRevLett.107.143003}. Finally, four PnA cycles with an evolution time of $10\,\text{s}$ are performed to determine the modified cyclotron frequency with highest precision.

After that, the ion in the PT is transported into one of the STs and the second ion is transported into the PT. With this ion the whole procedure is repeated in reverse order. This way, the time between the high-precision determination of the modified cyclotron frequencies is less than five minutes. Consequently, the jitter of the magnetic field during this time span contributes to the statistical fluctuations of each CFR. The double-dip technique and the PnA detection method are both used during the measurement cycle, which provides a crucial internal consistency check on the determination of $\nu_+$.

$\nu_-$ is determined at the beginning, middle and end of the whole measurement campaign, which lasted two months. The magnetron frequency is measured using the double-dip technique, which leads to an uncertainty of $100\,\text{mHz}$. This is absolutely sufficient to not limit the precision of the CFR.

\begin{flushleft}
\textbf{Statistical result}
\end{flushleft}

During the evolution time of the PnA method the ions are oscillating with a certain magnetron, axial and modified cyclotron amplitude. To get the rest mass, one has to take into account energy-dependent frequency shifts. This could be achieved via an independent energy calibration. However, we apply an extrapolation to zero excited modified cyclotron energy. To allow this, the modified cyclotron radius and therefore its energy is varied during the measurement campaign. While this extrapolation removes any effects depending on the excitation of the modified cyclotron motion, the effect of the thermal motion remains and has to be accounted for individually. Additionally, the slopes of the extrapolation provide a cross check for our independent energy calibration.

Over the course of the measurement campaign, in total 13 runs have been performed. A run consists on average of 30 cycles and in between runs, certain parameters like the excitation strength have been varied. For each different run $i$, we calculate the  $\text{CFR}=\frac{\nu_c \left(^{12}\text{C}^{6+}\right)}{\nu_c \left(\text{p}\right)}$ for every single cycle in this run and calculate the mean and the standard deviation of all cycles. This yields a CFR$_i$ with an uncertainty. The chi-square goodness-of-fit test as it is implemented in MATLAB~\cite{matlab} does not reject the null hypothesis that the input data for the single CFR$_i$ is normally distributed at a $5\%$ significance level.

\begin{figure}
\includegraphics[width=0.483\textwidth]{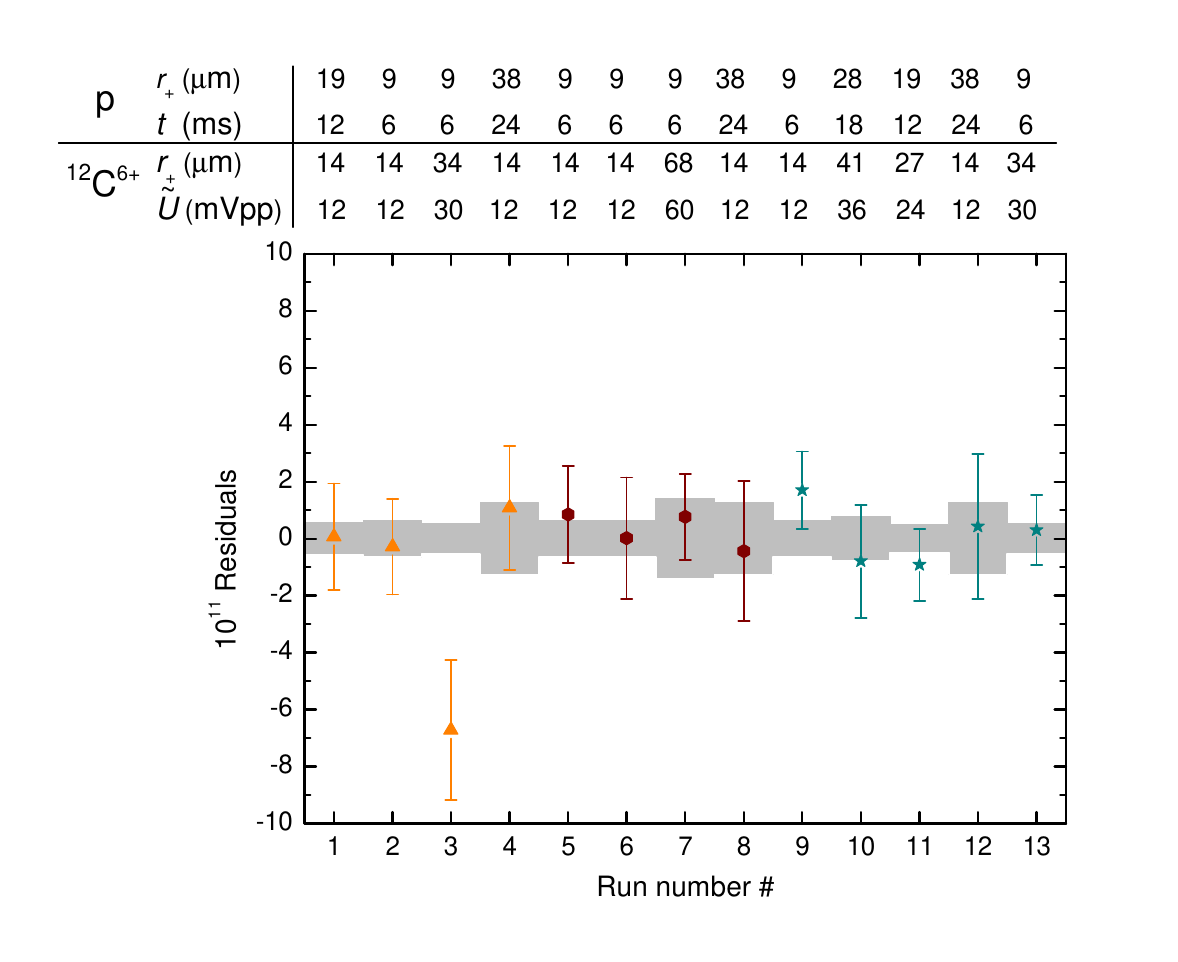} 
\caption{\label{resi_plot} Residual plot for the single CFRs for the different runs. For the proton the excitation amplitude $\tilde{U} = 40\,\text{mVpp}$ was constant during the whole measurement campaign, whereas for the carbon ion the excitation time $t=300\,\text{ms}$ was constant. The corresponding cyclotron radii during the PnA measurements are shown, too. The three different ion pairs are shown in different colors. The light gray area represents the one sigma prediction interval for each radius, based on the three parameter planar fit. For details see text.}
\end{figure}

We apply a three parameter (planar) fit with the offset $\text{CFR}_{\text{stat}}$ and the two excitation strengths of the proton and carbon ion as fit parameters $\left(S^+_{t,\tilde{U}} \left(\text{p}\right)~\text{and}~S^+_{t,\tilde{U}}\left(^{12}\text{C}^{6+}\right)\right)$. The residuals of the different runs together with the corresponding cyclotron radii $r_+^{\text{exc}}$ based on their individual excitation strengths and times for the different ion pairs during the whole measurement campaign are shown in Fig.~\ref{resi_plot}. 

The 13 CFR$_i$ are the input data for the planar fit. Altogether three different ion pairs are used to exclude systematic effects from undiscovered contaminant ions within the trap or the order of the ions. All the observed CFRs of each run are fitted with the following planar function including the three parameters CFR$_{\text{stat}}$, $a$ and $b$:
\begin{eqnarray}
\text{CFR}_i &=&\text{CFR}_{\text{stat}} + a \cdot  \left[ S^+_{t,\tilde{U},i} \left(\text{p}\right) \right]^2 + \nonumber \\
& & b \cdot \left[ S^+_{t,\tilde{U},i}\left(^{12}\text{C}^{6+}\right) \right]^2 \quad .
\end{eqnarray}

During the whole measurement campaign only the lengths of the first PnA pulses $t$ were varied for the proton to exclude systematic effects such as nonlinearities in the voltage output of the wave form generator or in the transfer function of the excitation lines. For the carbon ion, however, the amplitude had to be modified, to avoid excessively long excitation times. It is additionally possible to calculate the $B_2$ inhomogeneity based on the slopes $a$ and $b$ and the axial shift, see Eq.~(\ref{eq:spec}). This cross check yields a quadratic inhomogeneity which is in agreement with the previously determined one.

The fit parameter $\text{CFR}_{\text{stat}}$ gives the CFR extrapolated to zero excitation radius for the proton and the carbon ion:
\begin{equation}
\text{CFR}_{\text{stat}}=0.503\,776\,367\,640\,1(81) \quad .
\end{equation}
This is the statistical result without any systematic correction applied yet. The $\chi^2$ test yields $\chi_{\text{red}}^2=1.17$. The probability of a larger $\chi^2$ for ten degrees of freedom is $30\,\%$, which is a hint that the whole data is normally distributed and no superstatistical fluctuations occurred during the measurement. Since the data set consists of three different pairs of ions (with exchanged order in the trap tower), the absence of superstatistical fluctuations render effects like a dependence on the order of ions or electrons trapped in between the ions unlikely. Moreover, the reduced $\chi^2$ gives us confidence in our systematics model, because the CFR measurements are in good agreement at significantly different cyclotron radii. The statistical result slightly differs from the one reported in~\cite{PhysRevLett.119.033001} due to a minor flaw in our measurement program, discovered in the reanalysis.

The statistical uncertainty is dominated by magnetic field fluctuations during the time between the $T_{\text{evol}} =10\,\text{s}$ measurements of the PnA method for the two ions.
\begin{figure}
\includegraphics[width=0.48\textwidth]{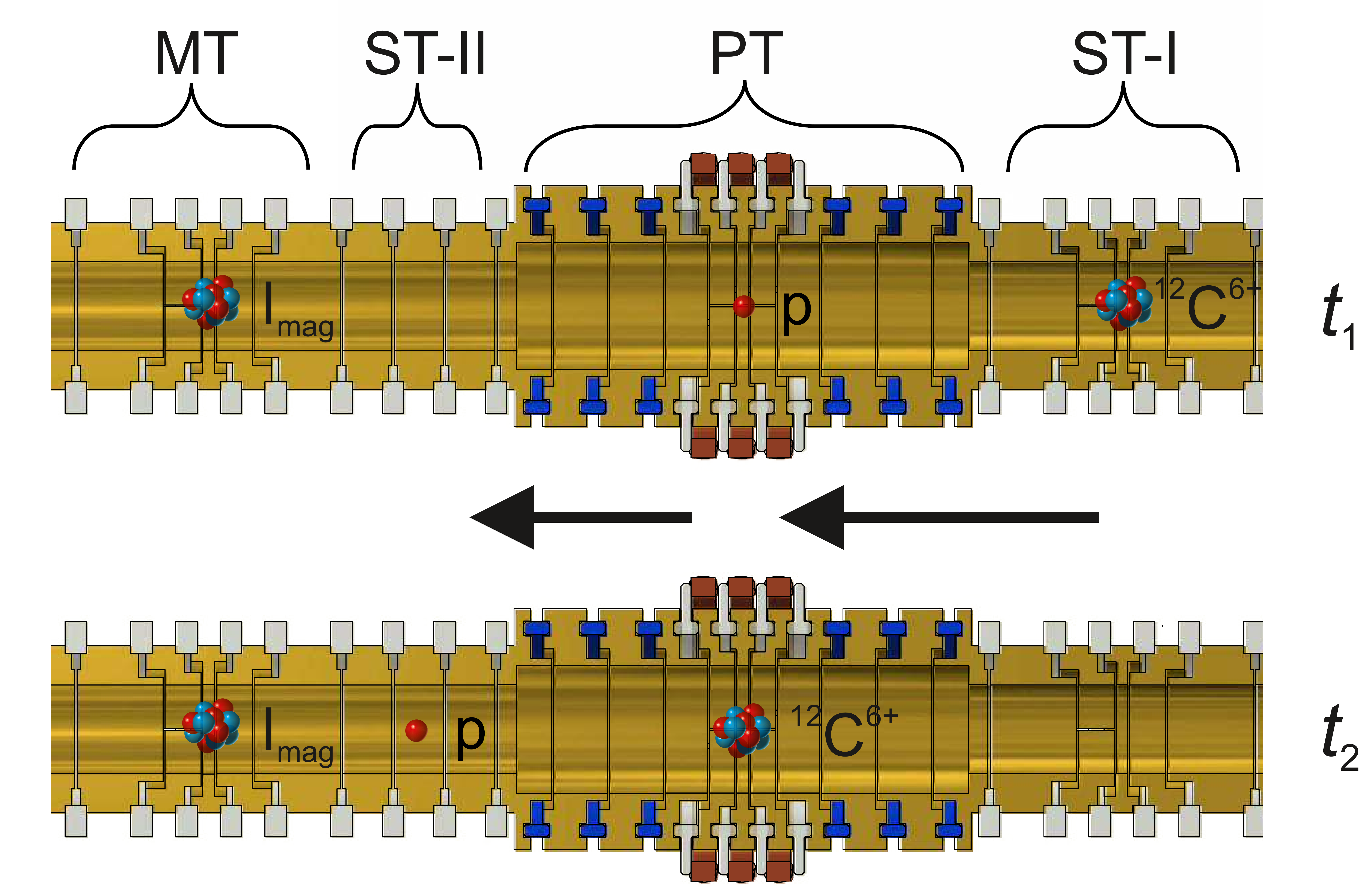} 
\caption{\label{sym_pna} Schematic view of the measurement principle of the simultaneous phase-sensitive method. For this method, three single ions are required, which are stored in different traps. During $t_1$ the CFR-1 of ion-1 (I$_1$), stored in the PT, and the magnetometer ion, stored in the MT, are measured simultaneously. In the second time step $t_2$ ion-II (I$_2$) is transported to the PT and I$_1$ is stored in the ST-II, followed by a CFR-2 measurement of I$_2$ and the magnetometer ion. The common-mode magnetic field fluctuations of the MT and PT are canceled to a large extent by combining these two CFRs.}
\end{figure}
The MT will allow a simultaneous phase-sensitive measurement of the CFR and thus helps to overcome the magnetic field fluctuations during the ion exchange in the PT. For such a measurement scheme a third ion $\text{I}_{\text{mag}}$ is stored in the MT and its cyclotron frequency $\nu_c\left(\text{I}_{\text{mag}}\right)$ is measured simultaneously with the ion stored in the PT, see Fig.~\ref{sym_pna}. This scheme is similar to the one which will be applied at the PENTATRAP experiment~\cite{Repp2012}. The simultaneous phase-sensitive measurement method will probably allow for much longer measurement times, since the magnetic field fluctuations are canceled to a large extent. For the PnA method this could potentially lead to a significantly simpler unwrapping of the phases for long evolution times.

Independently, it is possible to improve the magnetic field stability by pressure stabilization of the helium reservoirs of the magnet and the apparatus. The combined pressure dependence of $\nu_+$ for these reservoirs is preliminarily determined to be $\Delta \nu_+ \approx 80\,\text{mHz/mbar}$ at $\nu_+ \approx 29\,\text{MHz}$. The pressure can be stabilized to a few $\mu$bar, which can potentially increase magnetic field stability dramatically.

Another statistical limitation arises from the axial frequency stability. Up to now we identified two sources of jitter for the axial frequency, voltage fluctuations and fluctuations of the angle between the magnetic field lines and the trap axis~\cite{Marc2018}. The angle fluctuations affect both traps simultaneously, whereas the voltage jitter is uncorrelated to a very large extent, since the traps are connected to two independent voltage references. Equation~(\ref{magistabi}) holds as long as the jitter due to voltage fluctuations is small compared to the jitter of $\nu_+$ during $T_{\text{evol}}$. A relative voltage stability of better than $\delta U/U = 7\times 10^{-8}$ within five minutes results in $\delta \nu_z \left( ^{12}\text{C}^{6+}\right) = 15\,\text{mHz}$ and is required for a relative statistical uncertainty of $1 \times 10^{-11}$ for the CFR. To reach this goal, additional stabilization techniques are required in the future, since the present stability over five minutes corresponds to an effective voltage stability of $\delta U_{\text{fast}}/U \leq 2 \times 10^{-7}$ for the PT. Only an upper limit can be given, since it is up to now not possible to disentangle the different sources of the axial frequency fluctuations.

\begin{flushleft}
\textbf{Systematic shifts}
\end{flushleft}

The measured  $\text{CFR}_{\text{stat}}$ needs to be corrected for systematic shifts, which are summarized in Table~\ref{sys_un}.

\begin{table} 
\begin{center}
\caption{Overview of systematic contributions resulting in shifts of $\text{CFR}_{\text{stat}}$.}\label{sys_un}
\begin{ruledtabular}
\begin{tabular}{l c c}
\multirow{2}{*}{Effect} & $\frac{\text{CFR}_{\text{stat}}-\text{CFR}_{\text{cor}}}{\text{CFR}_{\text{stat}}}$  & Uncertainty \\
 &  ($10^{-12}$) &  ($10^{-12}$) \\
 \hline
Image charge & 91.0 & 4.6\\
Image current & $-1.9$ & 0.3 \\
Lineshape model\footnote{Between the three ion pairs this value varies slightly due to different $\nu_z$ in relation to $\nu_{\text{res}}$.} & 3.1 & 3.0 \\
$\nu_-$ determination & 0.0 & 0.6 \\
Residual &  & \\
magnetic & $-20.9$   & 27.4 \\
inhomogeneity &  &  \\
Residual &  & \\
electrostatic & $\ll 0.1$ & $\ll 0.1$\\
anharmonicity & & \\
Special relativity & $-8.9$ & 7.1 \\
\hline
Total & 59.5 & 28.8 \\
\end{tabular}
\end{ruledtabular}
\end{center}
\end{table}

For the proton mass campaign we optimized the tuning ratio so that $C_4=\left(0\pm 6.3\right) \times 10^{-6}$ and $C_6=\left(-6.8\pm 0.4 \right) \times 10^{-4}$. These small anharmonicities lead to systematic shifts smaller than $10^{-13}$ of CFR$_{\text{stat}}$. After the $m_{\text{p}}$ measurement the tuning ratio was further improved, which is presented in section~\ref{chap_MT}. 

The motional magnetic moment of the ions due to the modified cyclotron energy and the gradient of the magnetic field lead to a net force in axial direction. At the equilibrium position the force is compensated by the force of the electrostatic gradient. The uncompensated $B_1$-component leads to a relative systematic shift smaller than $10^{-12}$ of CFR$_{\text{stat}}$. 

The systematic shift caused by the $B_2$ component of the magnetic field together with the thermal energy of the ion mainly affects the modified cyclotron frequency of the proton due to its small mass.  For the carbon ion this shift is reduced by a factor of six. 

The uncertainty of the systematic shift caused by the $B_2$-component is the largest of all effects. To minimize this effect, it is favorable to compensate these magnetic inhomogeneities using shim coils and reduce the ion temperatures. Two superconducting shim coils have been wound to compensate the $B_1$ and $B_2$ magnetic field inhomogeneity components. These shim coils are mounted to the trap chamber. The technical details of the coils can bee found in Appendix~\ref{mag_shim_coils}.

The relativistic effect mostly impacts the proton, giving $\delta \nu_c /\nu_c \approx E_+ / (m c^2)=-75\,\text{ppt}$ for the proton during the PnA cycle with the smallest $r^{\text{exc}}_+$. Due to the extrapolation of the modified cyclotron excitation energy to zero, only the thermal energy of the ions leads to a systematic uncertainty.

The two cyclotron resonators were mounted to be used for further cooling the axial motion of the ions below ambient helium temperature by sideband coupling. With a reasonably fast cooling time constant, much faster compared to the heating rate, it is possible to reduce the axial temperatures to $T_z (\text{p}) \approx   13 \,\text{mK}$ and $T_z\left(^{12}\text{C}^{6+}\right) \approx   18 \,\text{mK}$ via sideband coupling of the modified cyclotron and axial mode to the corresponding cyclotron resonator at $4\,\text{K}$. At the moment, however, the $Q$-values of both cyclotron resonators are too low, and correspondingly $\tau_+$ is too large for cooling these modes effectively.

Besides the energy dependent shifts, several energy independent shifts also affect the CFR.
The induced image charges on the trap electrodes act back onto the motion of the ion. This is the so-called image charge shift of the modified cyclotron frequency and the magnetron frequency:
\begin{eqnarray}
\Delta \nu_{\pm} &=& \mp C_{\text{ICS}} \frac{m}{8 \pi \epsilon_0 r_0^3 B_0^2} \quad , \\
 \quad C_{\text{ICS}} &=& 1.97(10) \quad .
\end{eqnarray}
Here, $C_{\text{ICS}}$ is a coefficient that depends on the geometry of the trap and that can be calculated. The effect for $ \nu_z $ is very small and arises due to the slits of the electrodes, which break the axial mirror symmetry. The PT has a larger trap radius of $5\,\text{mm}$ compared to $3.5\,\text{mm}$ of the former \textit{g}-factor trap, which reduces the image charge shift by a factor of three. Still, this effect is responsible for the largest systematic shift. For $\nu_c\left(^{12}\text{C}^{6+}\right)$ this results in a relative shift of $99 \times 10^{-12}$, whereas for $\nu_c \left(\text{p}\right)$  it is  $8.3 \times 10^{-12}$. The value of $C_{\text{ICS}}$ and its corresponding relative uncertainty of $5\%$ is based on numerical simulations and depends on the trap design~\cite{Marc2018}. Recent simulations on the image charge effect, which are also experimentally tested, will further reduce the uncertainty of this systematic shift~\cite{Marc2018}. 

Besides the image charge of the ion, also its current acts back to the ion. The resonator impedance $Z_{LC}$ is a complex quantity. The imaginary part of the damping constant $\gamma$ shifts the frequency of the ion by $\Delta \nu = \nu^{\text{meas}} - \nu^{\text{ideal}}=-\text{Im}(\gamma)/(2 \pi)$~\cite{PhDNatarjan,Sven2012}, which is called image current shift or frequency pulling. Our lineshape model for the axial frequency already includes this shift and directly extracts $\nu_z^{\text{ideal}}$. For the proton mass measurement we detune the cyclotron resonators as far as possible by using the varactor diodes to lower this effect for the modified cyclotron frequency. $\nu_c\left(^{12}\text{C}^{6+}\right)$ is shifted by $-3 \times 10^{-12}$ and $\nu_c \left(\text{p}\right)$ by  $-1\times 10^{-12}$ relative to the measured frequency.

The axial frequency is determined for each ion during each cycle via the dip technique. There is an additional shift of the axial frequency due to an off-resonant position of the ion's frequency compared to the resonance frequency of the axial resonator, which is not included in our lineshape model. $\nu_c$ is shifted by $2.1 \times 10^{-12}$ for the carbon ion and $-1.0\times 10^{-12}$ for the proton. The exact shift changes due to different detunings of the ions compared to $\nu_{\text{res}}$ over the whole measurement campaign. 

$\nu_-$ is the smallest of all three eigenfrequencies of the ion and its uncertainty has therefore the smallest impact on $\nu_c$. The uncertainty of the magnetron frequency determination leads to an uncertainty of $ 0.1 \times 10^{-12}$ and $0.6 \times 10^{-12}$ on the respective $\nu_c$ for the proton and the carbon ion.

\begin{flushleft}
\textbf{Results}
\end{flushleft}

Applying all systematic shifts yields the following corrected CFR:
\begin{equation}
\text{CFR}_{\text{cor}}\vert_{\text{stat,sys}}~=~0.503\,776\,367\,670\,1 (81)(144) \quad .
\end{equation}
Applying this value and $m(^{12}\text{C}^{6+})$ in Eq.~(\ref{eq:3}) yields the atomic mass of the proton:
\begin{equation}
m_{\text{p}} = 1.007\,276\,466\,598(16)(29) \, \text{u} \quad ,
\end{equation}
with the relative uncertainty:
\begin{equation}
\frac{\delta m_{\text{p}}}{m_{\text{p}}}  = 3.3 \times 10^{-11} \quad . \label{res:prot}
\end{equation} 
The first bracket gives the statistical uncertainty, whereas the second one represents the total systematic uncertainty. The result is shifted by $1.5 \times 10^{-11}\,\text{u} \equiv 0.45 \,\sigma $ compared to the one reported in \cite{PhysRevLett.119.033001} due to the corrected motional temperatures of the particles. This result is a factor of three more precise compared to the CODATA value of 2014 and also shows a deviation of $3.1$ standard deviations to it. 

Applying this result to the fundamental constant $\mu = m_{\text{p}} / m_{\text{e}} $ leads to a two times more precise value compared to CODATA 2014:
\begin{equation}
\mu = 1\,836.152\,673\,374(78) \quad , 
\end{equation}
with the relative uncertainty:
\begin{equation}
 \frac{\delta \mu }{  \mu}  = 4.3 \times 10^{-11} \quad .
\end{equation}
Here, the total uncertainty is calculated by the square root of the quadratic sum of the electron and proton mass uncertainties, since the largest systematic uncertainties are different for both measurements. The systematic uncertainty of the proton mass is mainly given by the uncertainty of the quadratic magnetic field inhomogeneity, whereas the uncertainty of the electron mass is dominated by the uncertainty of the image charge effect. Therefore, no correlation of the uncertainties for the two measurements is assumed. This result is in good agreement with recent results determined by other experiments using different approaches~\cite{Hori610,Alighanbari18}.

The result for $m_{\text{p}}$ changes also the mass of the neutron, since it is determined via the mass of the deuteron subtracted by the deuteron binding energy and the mass of the proton~\cite{0026-1394-52-2-280}. The shift lies within the uncertainty of the neutron mass. Furthermore, $m_{\text{p}}$ also affects the puzzle of light ion masses and $R_{\infty}$ as discussed in section~\ref{intro_1} of this article. 

It is possible to determine the CFR$_{\text{DD}}$ only based on the measured double-dip frequencies instead of using the PnA results, since $\nu_+$ is also measured every cycle using the double-dip technique:
\begin{eqnarray}
\text{CFR}_{\text{DD}}\vert_{\text{stat,sys}}&=& 0.503\,776\,367\,68(3)(5) \quad , \\
m_{\text{p}} \left(\text{DD}\right) &=& 1.007\,276\,466\,61(6)(10)  \, \text{u}  \quad , 
\end{eqnarray}
with the relative uncertainty:
\begin{equation}
\frac{\delta m_{\text{p}}}{m_{\text{p}}} = 1.2 \times 10^{-10} \quad . 
\end{equation}
This extracted proton mass is in very good agreement to the one based on the PnA method, but is less precise by a factor of four.

\section{Determination of $m\left(^{16}\text{O}\right)$ } \label{mo_stat}

Due to the varactor diode of the axial resonator dedicated for $^{12}\text{C}^{6+}$, it was possible to shift the resonance frequency of the detection circuit $\nu_{\text{res}}$ from $\nu_z\left(^{12}\text{C}^{6+}\right) =525\,141\,\text{Hz}$ to $\nu_z\left(^{16}\text{O}^{8+}\right) =525\,216\,\text{Hz}$. With this adjustment we measured the CFR of oxygen $^{16}\text{O}^{8+}$ against the proton with the identical measurement cycle as presented above. Applying the corresponding Eq.~(\ref{eq:2}) for $^{16}\text{O}$ and its ionization energies~\cite{NIST_IP} results in: 
\begin{eqnarray}
m \left(^{16}\text{O}^{8+}\right) &=& 8 \frac{\nu_c\left(\text{p}\right)}{\nu_c\left(^{16}\text{O}^{8+}\right)} m\left(\text{p}\right) \\
 &=& \frac{8} { \text{CFR}\left(^{16}\text{O}^{8+}, \text{p} \right)}  m\left(\text{p}\right) \quad . \label{eq:ox}
\end{eqnarray}
Using the measured $\text{CFR}\left(^{16}\text{O}^{8+}, \text{p} \right) =0.503\,936\,558\,242(17)$  and applying all the corresponding systematic shifts yields:
\begin{equation}
m\left(^{16}\text{O}\right) = 15.994\,914\,619\,37 (54)(45)(51)\,\text{u} , 
\end{equation}
with the relative uncertainty:
\begin{equation}
\frac{\delta m\left(^{16}\text{O}\right)}{m\left(^{16}\text{O}\right)} = 5.4 \times 10^{-11}\quad .
\end{equation}
Again the first bracket represents the statistical uncertainty, whereas the second one is the total systematic uncertainty and the last one is caused by the uncertainty arising from the proton mass (Eq.~(\ref{res:prot})). The result is shifted by $1.3 \times 10^{-10}\,\text{u} \equiv 0.15 \,\sigma $ compared to the one reported in \cite{PhysRevLett.119.033001} due to the corrected motional temperatures of the particles. Our result is in very good agreement with the current literature value of the AME2016~\cite{1674-1137-41-3-030002,Wang_2017}:
\begin{equation}
m_{\text{AME2016}}\left(^{16}\text{O}\right)= 15.994\,914\,619\,60 (17) \, \text{u} \quad , 
\end{equation}
with the relative uncertainty:
\begin{equation}
 \frac{\delta m\left(^{16}\text{O}\right)}{m\left(^{16}\text{O}\right)} = 1.1 \times 10^{-11}\quad .
\end{equation}
The AME2016  $^{16}\text{O}$ mass value is mainly based on the CFR measurement by the group of Van Dyck Jr.~\cite{VanDyck2001,VANDYCK2006231}. Our result agrees within $0.3\,\sigma$ to the literature value and is the second most precise measurement of $m\left(^{16}\text{O}\right)$ so far.

We have performed a further consistency check by measuring the CFR of $^{12}\text{C}^{6+}$ and $^{12}\text{C}^{3+}$. For this CFR measurement $\nu_z \left(^{12}\text{C}^{6+}\right)$ was shifted to $\nu_{\text{res}}$ of the proton axial resonator: from $\nu_z\left(^{12}\text{C}^{6+}\right) =525\,141\,\text{Hz}$ to $\nu_z\left(^{12}\text{C}^{6+}\right) =739\,872\,\text{Hz}$. This was accomplished by doubling the applied voltages in the PT. This way it became possible to bring $\nu_z\left(^{12}\text{C}^{3+}\right) =528\,344\,\text{Hz}$ into resonance with the axial tank circuit originally designed for $\nu_z \left(^{12}\text{C}^{6+} \right)$. 

Since the atomic binding energies and the mass of the electron are well known, it is possible to calculate the expected CFR very precisely. Furthermore, the relative systematic uncertainties of the measurement are reduced to $8 \times 10^{-12}$ due to the identical nuclei. Using again a modified form of Eq.~(\ref{eq:3}), we insert $m\left(^{12}\text{C}^{3+}\right)$ and the measured CFR. The determined  $m\left(^{12}\text{C}^{6+}\right)$ has a relative uncertainty of $1.1 \times 10^{-10}$ and is in good agreement with the calculated one at a level of $0.2\,\sigma$. The uncertainty of this result is larger compared to the uncertainty of the proton mass mainly due to a shorter total measurement time. 

\section{Summary and outlook} \label{outlook}

In this paper the high-precision Penning-trap experiment $\text{LIONTRAP}$ is presented in detail. This includes the most harmonic Penning trap and for the first time four precisely tuned detection systems for one trap. Additionally, the proton's and oxygen's atomic mass measurements are discussed including a corrected value for both masses.

A further reduction of the statistical uncertainty might be accomplished by applying the simultaneous phase-sensitive measurements in the precision and magnetometer trap. This scheme leads to the cancellation of common-mode magnetic field fluctuations. Additional magnetic compensation coils have been produced and methods for reducing the temperature of the ions will be implemented in the future to further increase precision. Together with another measurement of the image charge effect this reduces the largest uncertainties of the $m_{\text{p}}$-measurement. With these upgrades we are aiming for ppt relative precision measurements of light atomic masses. 

In the next step we are focusing on the atomic mass of the deuteron. Together with a more precise determination of the deuteron binding energy and the measured proton mass, it is possible to improve the precision of the atomic mass of the neutron, too. Furthermore, this measurement can be another step towards a resolution of the puzzle of light ion masses.

\section*{Acknowledgements}
All authors from reference~\cite{PhysRevLett.119.033001} contribute and agree to the corrected proton mass value. We want to thank Andreas Mooser and Stefan Ulmer for the fruitful collaboration during the proton mass campaign. Furthermore, we like to thank Sven Junck, Jiamin Hou and Anke Kracke for their support in planning and building the experiment. We want to thank Marc Schuh for the calculation of the image charge shift in our trap.
This work was supported by the Max Planck Society as well as the International Max Planck Research Schools for Quantum Dynamics in Physics, Chemistry and Biology (IMPRS-QD) and for Precision Tests of Fundamental Symmetries (IMPRS-PTFS).

\appendix

\section{Determination of patch potentials} \label{patch_app}

In our trap the patch potentials are investigated via two different methods: First, the voltage of the varactor diode is changed leading to a shift in $\nu_{\text{res}}$. Then the ring voltage is tuned to bring the ion's axial frequency back into resonance with the detection circuit ($\nu_z \approx \nu_{\text{res}}$). One can determine the offset potentials by measuring both axial frequencies and voltages. In a second approach we measured the signal of $^{12}\text{C}^{6+}$ on the proton axial resonator at $U_{\text{R}} \approx -19.5\,\text{V}$, using a voltage doubler. The measurement of both voltages and frequencies for $^{12}\text{C}^{6+}$ yields the effective offset potentials $U_{\text{patch}}$ for the ring electrode in the following way:
\begin{eqnarray}
\nu_{z,1} \left(^{12}\text{C}^{6+}\right) & = & 525\,058.4\,\text{Hz,~} U_{\text{R,1}} = 9.809\,787\,\text{V} , \nonumber \\
& & \\
\nu_{z,2} \left(^{12}\text{C}^{6+}\right) & = & 739\,870.8\,\text{Hz,~} U_{\text{R,2}} = 19.479\,758\,\text{V}, \nonumber \\
& & \\
\nu_{z,1} &=& \nu_{z,2}  \sqrt{\frac{U_{\text{R,1}}+ U_{\text{patch}}}{U_{\text{R,2}}+ U_{\text{patch}}}} \quad , \\
U_{\text{patch}} &=& \frac{U_{\text{R,2}} \left(\frac{\nu_{z,1}}{\nu_{z,2}}\right)^2  -U_{\text{R,1}}}{1-\left(\frac{\nu_{z,1}}{\nu_{z,2}}\right)^2} \quad .
\end{eqnarray}
Assuming negligible leakage currents, the patch potentials are $U_{\text{patch}} < 10\,\text{mV}$ for the PT and thus more than an order of magnitude smaller than in the precision trap of the former \textit{g}-factor HCI experiment. Possible reasons are the more careful surface treatment during the production of the copper electrodes and during the silver and gold plating, resulting in a smaller surface roughness.

\section{Characterization of the seven-electrode cylindrical Penning trap} \label{harmonicty_trap}

To study the harmonicity of the electric potential the shift of the axial frequency due to an excitation of the magnetron motion is measured. The resulting shift of the axial frequency can be modeled using an even-order polynomial fit for the axial shift in dependence of the magnetron excitation strength $S^-_{t,\tilde{U}}$:
\begin{equation}
\Delta \nu_z = \sum_{i=1}^{\infty} p_i  \left(S^-_{t,\tilde{U}}\right)^{2i}  \quad ,
\end{equation}
where $S^-_{t,\tilde{U}}$ is the product of the pulse length $t_{\text{exc}}$ and the applied rf pulse amplitude $\tilde{U}_{\text{exc}}$ of the applied sinusoidal dipole excitation at the magnetron frequency. Neglecting insignificant terms with mixed amplitudes, e.g. $r_-^2 \cdot r_+^2$ and $r_-^2 \cdot z_0^2$ for $C_6$, leads to the expressions for the first even harmonicity coefficients $C_i$ ($i>2$):
\begin{eqnarray}
C_4 &=& - \frac{2}{3} \frac{C_2 d_{\text{char}}^2}{ \nu_z \kappa_-^2} p_1  \label{C4} \\
&=& E_4 + \frac{U_{\text{C1}}}{U_{\text{R}}} D_{4,1}  + \frac{U_{\text{C2}}}{U_{R}} D_{4,2} \quad ,\label{D-42}\\
C_6 &=& \frac{16}{45} \frac{C_2 d_{\text{char}}^4}{ \nu_z \kappa_-^4} p_2  \\
&=& E_6 + \frac{U_{\text{C1}}}{U_{\text{R}}} D_{6,1}  + \frac{U_{\text{C2}}}{U_{R}} D_{6,2}  \quad , \\
C_8 &=&  - \frac{8}{35}  \frac{C_2 d_{\text{char}}^6}{ \nu_z \kappa_-^6} p_3 \\
&=& E_8 + \frac{U_{\text{C1}}}{U_{\text{R}}} D_{8,1}  + \frac{U_{\text{C2}}}{U_{R}} D_{8,2}\quad , \\
C_{10} &=& \frac{256}{1575}  \frac{C_2 d_{\text{char}}^8 }{\nu_z \kappa_-^8} p_4  \\
&=& E_{10} + \frac{U_{\text{C1}}}{U_{\text{R}}} D_{10,1}  + \frac{U_{\text{C2}}}{U_{R}} D_{10,2}\quad , \\
C_{12} &=& - \frac{256}{2079}  \frac{C_2 d_{\text{char}}^{10}}{ \nu_z \kappa_-^{10}} p_5  \\
 &=& E_{12} + \frac{U_{\text{C1}}}{U_{\text{R}}} D_{12,1}  + \frac{U_{\text{C2}}}{U_{R}} D_{12,2}\quad .
\end{eqnarray}
Here, $\kappa_-$ is the proportionality constant between the magnetron excitation strength and the excited magnetron radius $r^{\text{exc}}_- = \kappa_- \cdot S^-_{t,\tilde{U}}$. The $D_{i,j}$ slope-coefficients can be determined from numeric simulations. Eq.~(\ref{C4}) together with Eq.~(\ref{D-42}) gives:
\begin{eqnarray}
p_1 &=& - \frac{3}{2} \frac{ \nu_z \kappa_-^2 C_4}{d_{\text{char}}^{2} C_2 }  \\
&=& - \frac{3}{2}  \frac{\nu_z \kappa_-^2 \left(E_4 + \frac{U_{\text{C1}}}{U_{R}} D_{4,1}  + \frac{U_{\text{C2}}}{U_{\text{R}}} D_{4,2} \right)}{d_{\text{char}}^{2} C_2}  .
\end{eqnarray}
The determination of the $p_1$-coefficients for different voltages of the second correction electrodes $U_{\text{C2}}$ with constant $U_{\text{R}}$ and constant $U_{\text{C1}}$ yields the following slope $m_{p_1}$:
\begin{equation}
m_{p_1} = \frac{\partial p_1}{\partial U_{\text{C2}}} = -\frac{3}{2} \frac{\nu_z \kappa_-^2 }{d_{\text{char}}^{2} C_2} \frac{D_{4,2}}{U_{\text{R}}} \quad .
\end{equation}
Finally, $\kappa_-$ can be determined from:
\begin{equation}
\kappa_- = \sqrt{-\frac{2 m_{p_1} U_{\text{R}} C_2 d^2_{\text{char}}}{3 \nu_z D_{4,2}}} \quad . \label{kappa}
\end{equation}
In our study, $D_{4,2}$ is taken from simulations, which yields $\kappa_-=6.425 \times 10^{-6}\,\text{m}/\left(\text{V}_{\text{pp}}\cdot \text{cycle}\right)=3.065 \times 10^{-2}\,\text{m}/\left(\text{V}_{\text{pp}}\cdot \text{s}\right)$. 
We measure the $D_{i,j}$ in the experiment to confirm the theoretical calculation, since they are rather robust against manufacturing uncertainties and patch potentials, see Table~\ref{tuningratio}.

\begin{table} 
\begin{center}
\caption{Comparison of the predicted and measured even coefficients $D_{i,j}$ of the electric potential. The uncertainties of the theoretical values arise from simulations by the variation of the lengths of the electrodes and distances between the electrodes by $\pm 20\,\mu\text{m}$. This study is performed with a $^{12}\text{C}^{6+}$ ion.}\label{tuningratio}
\begin{ruledtabular}
\begin{tabular}{c c c  c}
Coefficent & Theory  & Experiment & Deviation ($\%$) \\
 \hline
$D_{2,1}$ & -0.805(7) &  -0.816(1) & 1.4(9)	\\ 
$D_{2,2}$ & 	0.941(8) & 0.934(1) & 0.7(9) \\
$D_{4,1}$ & -0.105(19) & -0.067(3) & 36(18) \\
$D_{4,2}$\footnote{$D_{4,2}$ is used for the determination of $\kappa_-$, since the theory as well as the experimental results have small uncertainties.} &  -0.840(9) & -0.840(4) & 0.0(1.2)\\
$D_{6,1}$ & 0.997(30) &  1.051(41) & 5.4(5.1) \\
$D_{6,2}$ & -0.043(11)& 0.017(73) & 60(172)
\end{tabular}
\end{ruledtabular}
\end{center}
\end{table}

The slopes predicted by the theory are in agreement with the ones determined in the experiment. The uncertainties of the theory values are mainly caused by the manufacturing uncertainty of the trap electrodes, assuming an uncertainty of $\pm 20\,\mu$m. The increased uncertainty of the experimental values for higher-order coefficients reflects the larger excitation radius of the ion, resulting in an increasing impact of the finite ion temperature in combination with the lower-order trap anharmonicities.

\section{Magnetic field characterization} \label{mag_app}

\begin{figure}
\includegraphics[width=0.4\textwidth]{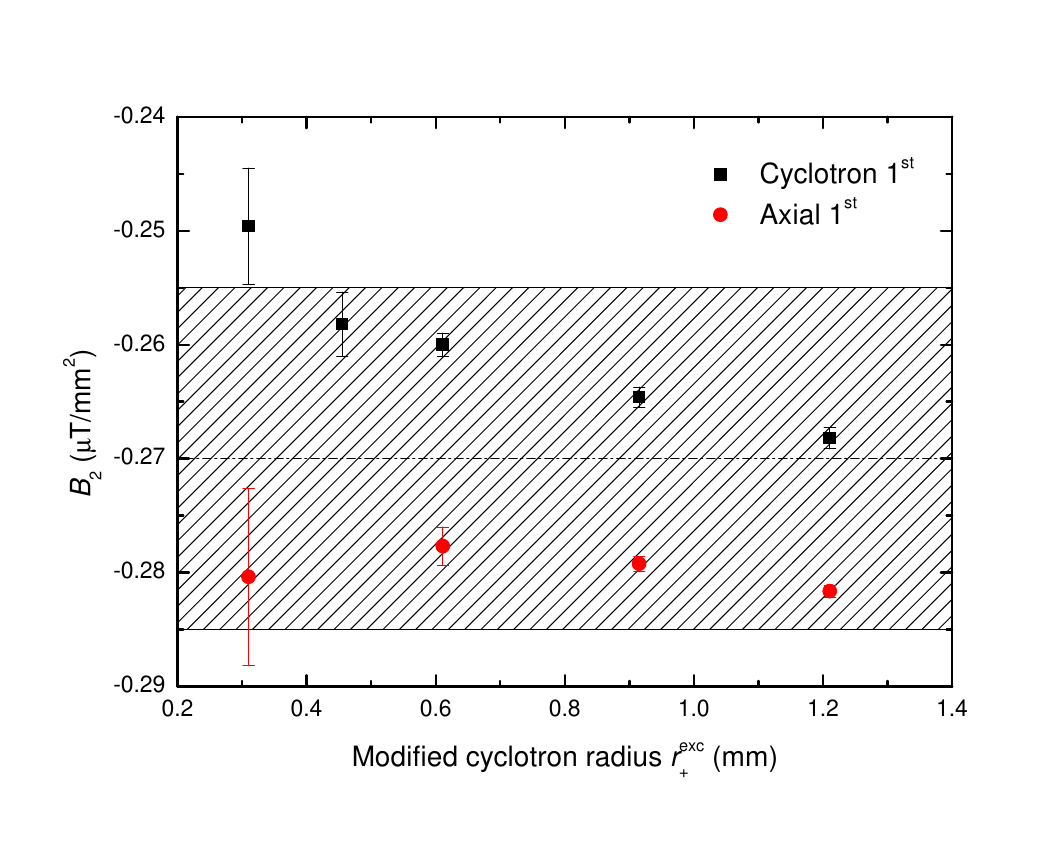}
\caption{\label{B2_1} Resulting $B_2$ for different excited modified cyclotron radii. The black data points are determined by a subsequent measurement of the modified cyclotron frequency shift via a direct peak detection on the corresponding cyclotron resonator followed by the determination of the axial frequency shift via the axial resonator. For the red data points the measurement order was inverted. The shaded area indicates the determined value: $B_2=-0.270\,(15)\,\mu\text{T/mm}^2$. For details see text.}
\end{figure}

In absence of electric field anharmonicities, shifts in the axial and modified cyclotron frequency occur due to $B_2$ and special relativity~\cite{RevModPhys.58.233,KETTER20141}:
\begin{eqnarray}
\Delta \nu_+ &=& - \frac{\left( 2 \pi \right)^2}{2} \frac{\left(\nu_+ - \nu_- \right)\nu_+^2}{c^2}  \left(r^{\text{exc}}_+\right)^2  \nonumber \\
& &  - \frac{1}{2}\frac{B_2 \left(\nu_+ + \nu_- \right)\nu_+}{ B_0 \left(\nu_+ - \nu_-\right)}  \left(r^{\text{exc}}_+\right)^2 \quad , \label {eq:spec} \\
\Delta \nu_z &=&  -\frac{\left( 2 \pi \right)^2}{4} \frac{\left(\nu_+ - \nu_- \right)\nu_+ \nu_z}{c^2}  \left(r^{\text{exc}}_+\right)^2   \nonumber \\
& & +  \frac{1}{4}\frac{B_2 \left(\nu_+ + \nu_- \right)\nu_z}{B_0 \nu_-}  \left(r^{\text{exc}}_+\right)^2 \quad , \label{eq:24}
\end{eqnarray}  
where $r^{\text{exc}}_+ = \kappa_+ \cdot S^+_{t,\tilde{U}}$. $S^+_{t,\tilde{U}}$ is the product of the duration and amplitude of the applied dipole modified cyclotron frequency excitation. The measurements of these two frequency shifts are required to determine $B_2$ and simultaneously the excitation parameter of the modified cyclotron mode $\kappa_+$. The result for different excited modified cyclotron radii is shown in Fig.~\ref{B2_1}. 
The modified cyclotron frequency shift is detected via a peak detection on the corresponding cyclotron resonator and the axial frequency shift is determined via a dip measurement. During the measurement time the ion's axial and modified cyclotron motion are resistively cooled by their corresponding axial and modified cyclotron tank circuits. However, the modified cyclotron frequency shift and the axial frequency shift are not determined simultaneously. If the axial dip is detected first, the modified cyclotron motion is resistively cooled during this time and therefore the afterwards detected cyclotron frequency shift gets smaller. The same occurs with the axial frequency shift if the modified cyclotron frequency shift is determined in the beginning. Therefore, the two methods lead to slightly different values of $B_2$. Simulations confirm the observed measured data and we can determine $B_2=-0.270(15)\,\mu\text{T/mm$^2$}$. A simultaneous measurement of the dip and the peak would result in a consistent value for $B_2$. But this was not possible in our case, since the modified cyclotron tank circuit needs to be detuned to prevent the thermalization of the modified cyclotron mode and thus resolve the axial frequency shift.

\begin{figure}
\includegraphics[width=0.4\textwidth]{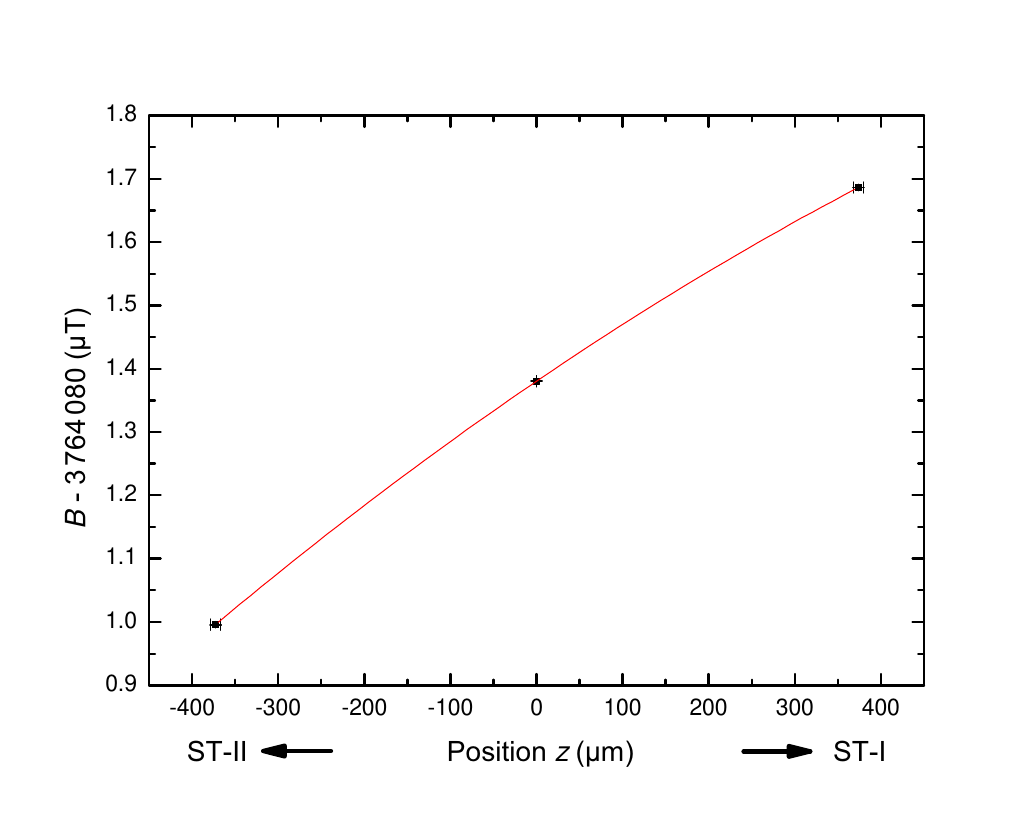}
\caption{\label{B2_2m} $B_2$ measurement by applying asymmetric trapping potentials. The $z$-position was determined based on calculated potentials, whereas the difference in the magnetic field is determined from the cyclotron frequency at each position. The red line is a second order polynomial fit. For details see text.}
\end{figure}

A second method to determine $B_1$ and $B_2$ is to map the $B$-field with measurements of $\nu_c$ at different positions along the $z$-axis of the trap. This is achieved by applying asymmetric potentials to the electrodes. The position of the ion is shifted by approximately $\pm 370\,\mu\text{m}$ along the $z$-axis. At each position the free cyclotron frequency was determined. The quadratic fit of Fig.~\ref{B2_2m} (red line) results in $B_2=-0.286(12)\,\mu\text{T/mm$^2$}$ and  $B_1=0.925(15)\,\mu\text{T/mm}$. The result for $B_2$ is in excellent agreement with the one mentioned before. The determined $B_2$ of the second attempt is the one in axial direction, whereas the first one is the $B_2$ in radial direction. 

\section{Ion temperature determination} \label{ion_temp_app}

We determine the axial temperature in two different ways. First by applying a setting with a large $C_4$. This shifts the frequency of the ion, broadens the width of the dip and reduces its depth. The lineshape $u_{\text{n}}^{\text{dip}}\left(\nu \right)$ of a regular dip signal with $C_4=0$ is measured in units of dBV$_{\text{rms}}$ and can be described by~\cite{Sven2012}:
\begin{widetext}
\begin{eqnarray} 
 u_{\text{n}}^{\text{dip}}\left(\nu | \nu_z , A, \tau ,Q,\nu_{\text{res}},u_{\text{n}}^{\text{off}},\kappa_{\text{det}}\right)  &=& 10 \log_{10}  \left(A  \frac{\text{Re} \left(Z_{\text{tot}} \left( \nu | \nu_z , A, \tau ,Q, \nu_{\text{res}},R_{\text{p}} \right) \right)}{R_{\text{p}}}  + u_{\text{n}}^{\text{off}} + \kappa_{\text{det}} \left(\nu - \nu_{\text{res}} \right) \right) \, (\text{dBV}_{\text{rms}}) , \nonumber \\ 
 \\
\text{Re} \left(Z_{\text{tot}} \right) &=& R_{\text{p}} \frac{ \left( \nu_{\text{res}} \nu \left( \nu^2-\nu_z^2\right) \right)^2}{\left(\nu_{\text{res}} \nu \left(\nu^2-\nu_z^2 \right)\right)^2 + \left( Q \left( \nu^2+\nu_{\text{res}}^2\right)\left( \nu^2-\nu_z^2\right)-\nu_{\text{res}}\nu^2/\left(2 \pi \tau\right)\right)^2} \quad ,
\end{eqnarray}
\end{widetext}
where $ Z_{\text{tot}}$ is the total impedance of the detection system including the thermalized ion signal. $\nu$ is the frequency in the Fourier spectrum and $u_{\text{n}}^{\text{off}}$ is the additional thermal noise of the cryogenic amplifier. $A$ is an amplification factor, which is determined at the optimum tuning ratio ($C_4=C_6=C_8=0$). The slope $\kappa_{\text{det}}$ describes the linear frequency dependence of the transfer function of the complete detection system. The parameters $ \nu_{\text{res}}, Q$, $u_{\text{n}}^{\text{off}}$ and $\kappa_{\text{det}}$ are determined via a fit of the noise spectrum of the resonator without the ion. They are fixed for the fit of the ion dip signal.

\begin{figure}
\includegraphics[width=0.48\textwidth]{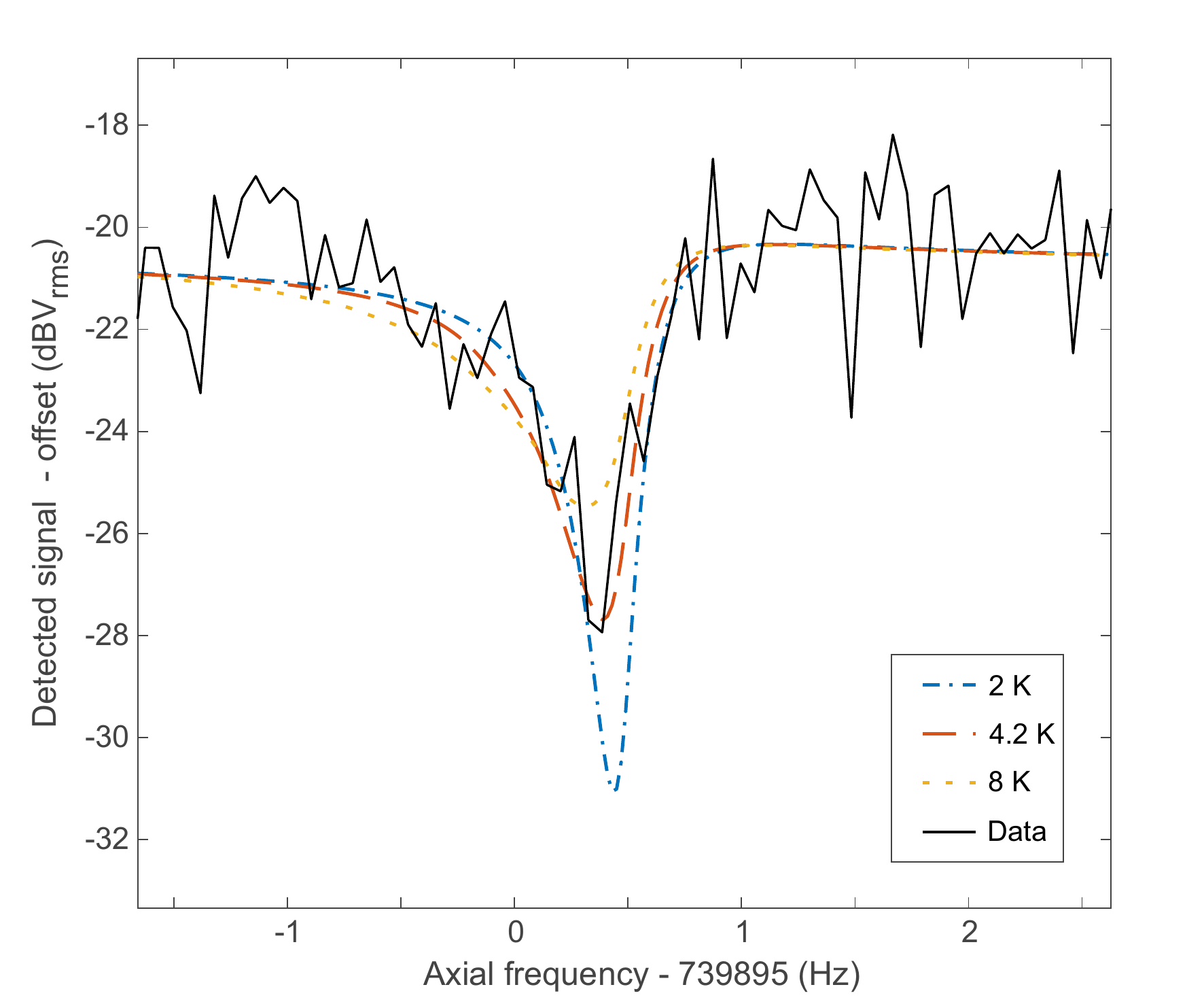}
\caption{\label{Temp_I} Axial temperature determination of the proton by studying the distortion of the dip signal in an anharmonic electric trapping potential. For details see text.}
\end{figure}

The anharmonicity coefficient $C_4$ modifies $\nu_z \rightarrow \nu_z' = \nu_z \left(1+3/2 \cdot C_4/(C_2 d_{\text{char}}^2) \cdot E_z/(4 \pi^2 m \nu_z^2) \right) $~\cite{KETTER20141}. $E_z$ varies on time-scales of the axial cooling time constant, while the ion is in equilibrium with the tank circuit. The final lineshape is expressed as a convolution of the regular dip lineshape with the thermal Boltzmann distribution of the axial energy:
\begin{widetext}
\begin{equation} 
u_{\text{n}}^{\text{dip,}C_4} =  10 \log_{10}  \left( \frac{1}{k_{\text{B}} T_z} \int_0^{\infty} e^{-\frac{E_z}{k_{\text{B}} T_z}}   u_{\text{n}}^{\text{dip}} \left(\nu | \nu_z' , A, \tau ,Q,\nu_{\text{res}},u_{\text{n}}^{\text{off}},\kappa_{\text{det}}\right)  dE_z + u_{\text{n}}^{\text{off}} \right) \, (\text{dBV}_{\text{rms}}) \quad .
\end{equation}
\end{widetext}
Using the dip signature at large $C_4$, the axial temperature of the proton is determined to $T_z =4(2)\,\text{K}$, see Fig.~\ref{Temp_I}. However, with this method it is not possible to determine the axial temperature with applied negative feedback, because at strong feedback the SNR of the noise spectrum of the resonator is very small and therefore the dip is hard to characterize.

For this reason another method is used, which is based on the study of axial frequency jitter because of the Boltzmann distribution of the modified cyclotron energy. Here, it is possible to determine the axial temperature via the standard deviation (jitter) of the axial frequency due to the combination of the quadratic magnetic field inhomogeneity and the thermal distribution of the modified cyclotron mode, see Eq.~(\ref{eq:24}). 

\begin{figure}
\includegraphics[width=0.4\textwidth]{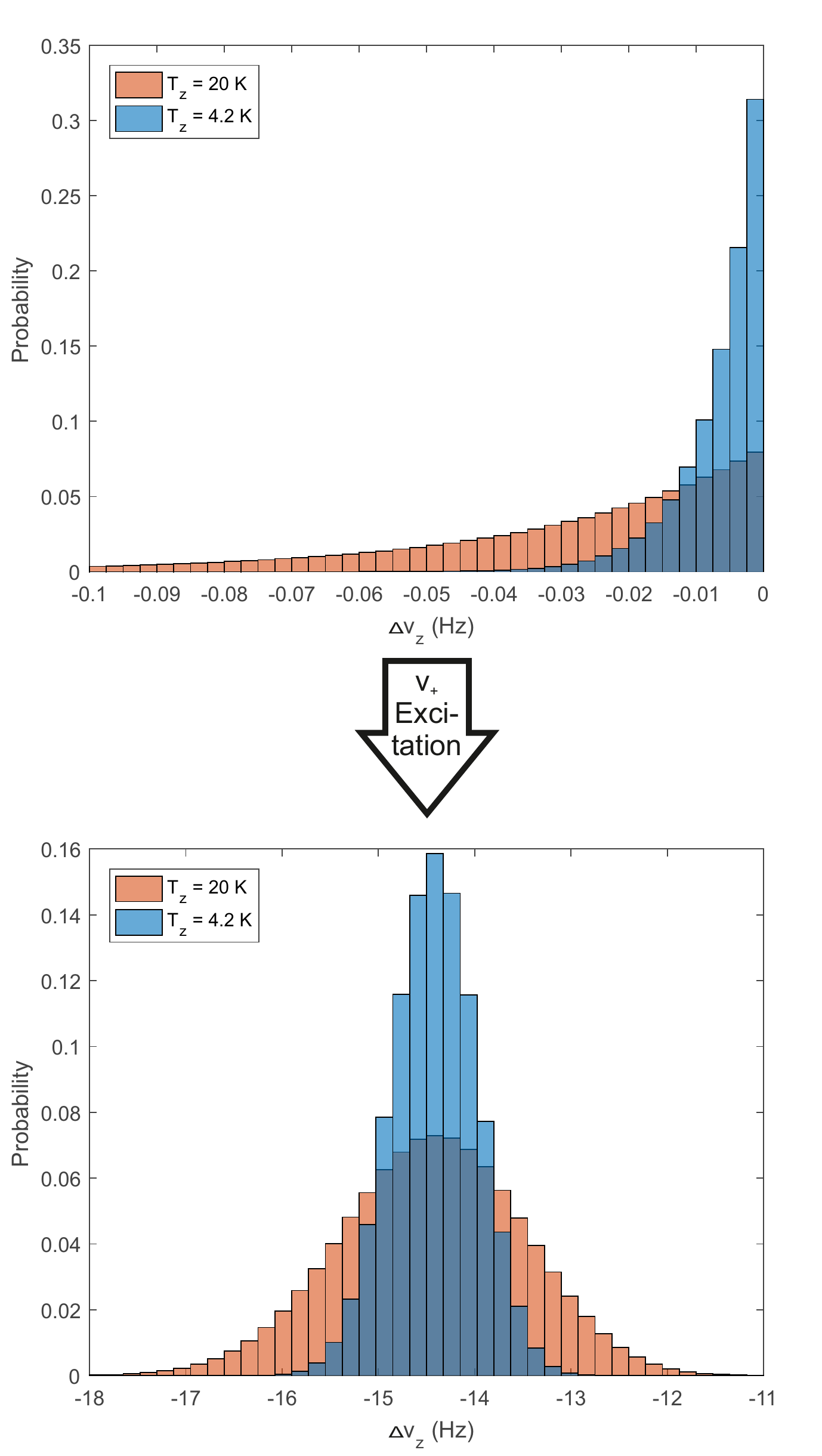}
\caption{\label{temp_fin} Simulated distributions of the axial frequency of the proton due to the thermal distribution of the modified cyclotron mode together with $B_2$, in red for $20\,\text{K}$ and in blue for $4.2\,\text{K}$. Upper plot: only thermalized distribution in $\nu_+$. Lower plot: excited thermal distribution ($r_+^{\text{exc}} \approx 300\,\mu\text{m}$).}
\end{figure}

For the proton with an axial temperature of $4.2\,\text{K}$ and sideband cooled eigenmotions ($\nu_+$, $\nu_-$) the jitter of $\nu_z$ due to $B_2$ and special relativity is around $7\,\text{mHz}$. For a lower temperature the fluctuations of the measured axial frequencies decrease. However, these small fluctuations are hard to resolve because of the resolution of the axial dip due to the trapping voltage. The voltage fluctuations occurring during $3\,\text{min}$ averaging time of the dip lead to a stability of $35\,\text{mHz}$ for $\nu_z$ for the proton. Therefore, the jitter of the axial frequency is determined after an excitation of the modified cyclotron motion to a radius of $ < r_+^{\text{exc}} > \approx 300\,\mu$m compared to a  thermalized radius of $ <r_+^{\text{therm}}> \approx 6\,\mu$m at $4.2\,\text{K}$. There, the jitter is $950\,\text{mHz}$ for a $20\,\text{K}$ cold proton and $440\,\text{mHz}$ for a $4.2\,\text{K}$ cold proton, due to the $B_2$ in our precision trap, which can be resolved, see Fig.~\ref{temp_fin}. A similar result can be achieved by exciting the magnetron motion in an electrostatic potential with an artificially large $C_4$ component. 

The finally observed jitter of the axial frequency is caused by voltage and thermal fluctuations described above. During the analysis for the $m_{\text{p}}$ paper~\cite{PhysRevLett.119.033001}, the jitter caused by voltage fluctuations was overestimated, leading to slightly lower axial temperatures. During the reanalysis of the complete data, this flaw was corrected. The result for the proton mass is mostly affected by the value for the carbon axial temperature with electronic feedback. This is discussed in section~\ref{mp_stat}.

Without using feedback for cooling, the axial temperature of the proton is determined to be $T_z(\text{p}) = 3.4(1.0)\,\text{K}$.
The axial temperature with applied feedback is $T_z(\text{p})=1.5(1.0)\,\text{K}$. The magnetron and cyclotron motions for the proton are cooled via sideband coupling to the axial tank circuit and additional feedback to $T_-(\text{p})= -10(6)\,\text{mK}$ and $T_+(\text{p})=116(78)\,\text{K}$, respectively. The eigenmotions of the carbon ion are cooled as well with negative feedback to $T_z=4.5(1.4)\,\text{K}$.

\section{Systematic shift of the lineshape model} \label{dip_shift}

\begin{figure}
\includegraphics[width=0.48\textwidth]{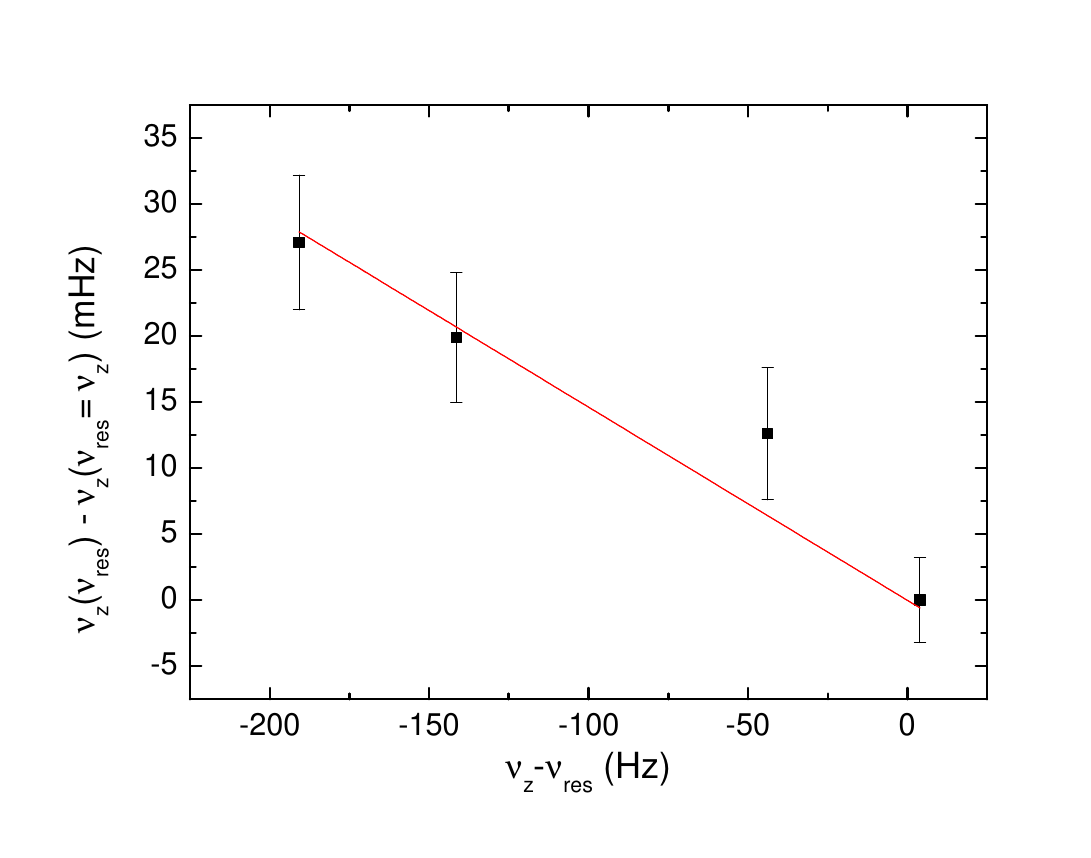} 
\caption{\label{varac} Axial frequency $\nu_z\left(^{12}\text{C}^{6+}\right)$ determined by the fit of our lineshape model for different resonance frequencies of the carbon detection circuit ($\nu_{\text{res}}$). Every other parameter stays constant during these measurements. The best fit yields a slope of $m_{\text{var}} = -1.46(4) \times 10^{-4}$. The single spectra are recorded with a frequency resolution of $1/32\,\text{Hz}$ and an acquisition time of $205\,\text{s}$.}
\end{figure}

For the first time, we were able to study our lineshape model at fixed voltage settings and different detunings of the carbon ion's $\nu_z$ compared to $\nu_{\text{res}}$ of the carbon resonator, see Fig.~\ref{varac}.

This investigation gives access to an additional shift of $\nu_z$ due to a slight off-resonant position of the ion's frequency compared to the resonance frequency of the axial resonator, which is not included in our lineshape model. Simulations confirm the measured trend by assuming voltage fluctuations of $\delta U_{\text{fast}} / U \approx 2 \times 10^{-7}$. They lead to a broadened dip, including a reduced dip-depth. 

Due to the asymmetric nature of the dip when detuned from the center of the resonator, this results in an axial frequency shift in dependence of the resonance frequency of the detection system when using our lineshape model. Dips of the proton axial frequency have a jitter of approximately $38\,\text{mHz}$ when comparing successive dip measurements with an averaging time of $200\,\text{s}$, which corresponds to a voltage fluctuations of $\delta U_{\text{slow}}  / U \approx 1 \times 10^{-7}$. Accordingly, there is probably a source of axial frequency fluctuations with a factor of two higher amplitude and at least a factor of two higher frequency compared to $\delta U_{\text{slow}}$.

\section{Magnetic field compensation coils} \label{mag_shim_coils}

\begin{figure}
\includegraphics[width=0.35\textwidth]{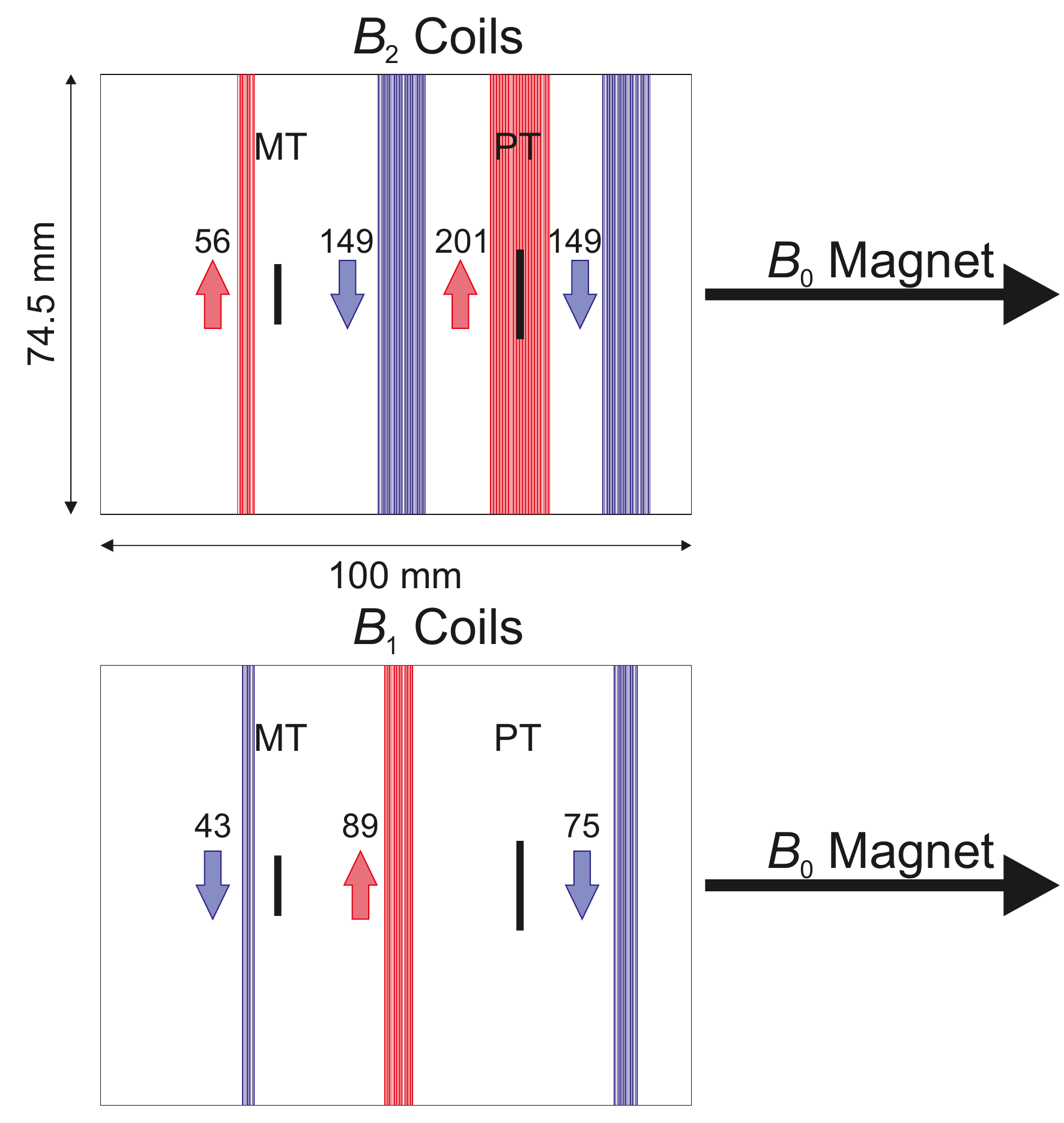} 
\caption{\label{b-shim} Geometry of the two compensation coils. The coils have been wound on an OFHC-copper cylinder, which is placed around the trap chamber. The color shows the winding sense and the number of windings is indicated. Both compensation coils are wound on the same body one above the other. Additionally, the position of the PT and MT is marked. The orientation of the magnetic field generated by the superconducting magnet itself is indicated by the black arrows.}
\end{figure}

\begin{figure}
\includegraphics[width=0.4\textwidth]{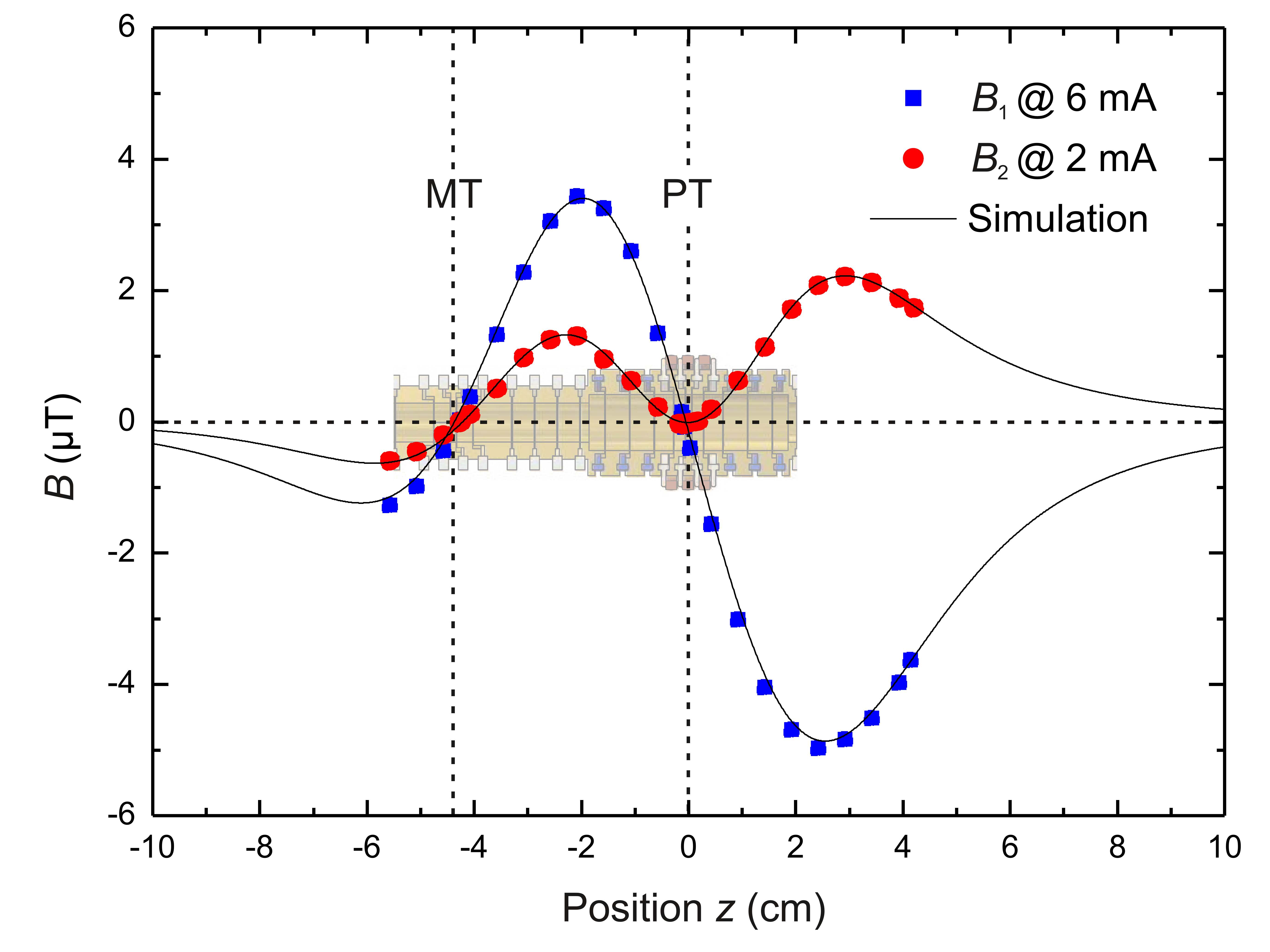} 
\caption{\label{shimcoil} Illustration of the generated magnetic field of the two compensation coils along the $z$-axis. The measured field is in excellent agreement with the simulation. In the background the position of the trap tower is shown. $z = 0$~cm is the position of the center of the PT. There, the generated magnetic field is close to zero to avoid an additional $B_0$. The second zero crossing is at the center of the magnetometer trap to also limit the magnetic field offset there. The generated $B_1$ and $B_2$ are approximately $-50\,\text{nT}/(\text{mm}\cdot\text{mA})$ and $3.5\,\text{nT}/(\text{mm}^2\cdot\text{mA})$, respectively. The measurement was done at room temperature outside of the magnet with a Hall effect sensor.}
\end{figure}

The shim coils consist of niobium-titanium wire with a total length of around $170\,\text{~m}$ and a diameter of $75\,\mu$m, see Fig.~\ref{b-shim}. The coils have been wound on an OFHC-copper cylinder, which is placed around the trap chamber. A potential magnetic field offset generated by the shim coils can fluctuate due to current instabilities of the supply. This fluctuating offset can potentially limit the stability of the overall magnetic field at the PT and MT. Accordingly, the generated $B_0$ at the positions of the ions in the MT and PT should be zero. The generated magnetic field is shown in Fig.~\ref{shimcoil}.

With a current of $34\,\text{mA}$ it was possible to generate a $B_1=1.6(1)\,\mu\text{T/mm}$ at the center of the PT, which compensates the residual $B_1$ of the magnet. Unfortunately, the superconductivity of the $B_2$ coil breaks down at a current of $16\,\text{mA}$. The generated $B_2=0.05(1)\,\mu\text{T/mm}^2$ at this current is a factor of five too small to completely compensate the $B_2$ in the PT. The most likely explanation for this is an inappropriate thermal coupling of the current supply lines, which heats up the coil. The applied power generates heat due to the resistance of the supply lines. Therefore, we did not use the compensation coil for the measurement campaign. The required power was reduced by closing the superconducting coils. Once loaded, the generated magnetic field of the compensation coils stays constant over time without a power connection. This will allow a sufficient compensation of the magnetic inhomogeneities for future measurements.

\bibliography{LIONTRAP_sources}
\end{document}